\documentclass[sigplan,screen]{acmart}  

\copyrightyear{2026}
\acmYear{2026}
\setcopyright{cc}
\setcctype{by}
\acmConference[ASPLOS '26]{Proceedings of the 31st ACM International Conference on Architectural Support for Programming Languages and Operating Systems, Volume 2}{March 22--26, 2026}{Pittsburgh, PA, USA}
\acmBooktitle{Proceedings of the 31st ACM International Conference on Architectural Support for Programming Languages and Operating Systems, Volume 2 (ASPLOS '26), March 22--26, 2026, Pittsburgh, PA, USA}
\acmPrice{}
\acmDOI{10.1145/3779212.3790154}
\acmISBN{979-8-4007-2359-9/2026/03}



\usepackage{tikz}
\usepackage{amsmath}
\usepackage{xurl}
\usepackage[shortlabels]{enumitem}
\usepackage[T1]{fontenc}
\usepackage{booktabs}
\usepackage{makecell}
\usepackage{pifont}
\usepackage{float}
\usepackage{subfig}
\usepackage{graphicx}
\usepackage[font={small}]{caption}
\usepackage{algorithm}
\usepackage{algpseudocode}
\usepackage[htt]{hyphenat}  
\usepackage{balance}

\definecolor{backgroundgray}{gray}{0.95}

\setlist[0]{leftmargin=1.5em}

\newcommand{\parasep}{\vspace{0.5ex}}
\newcommand{\heading}[1]{\parasep\noindent\textbf{#1.}~}

\newcommand{\ie}{i.e., }
\newcommand{\eg}{e.g., }

\newcommand{\figref}[1]{\mbox{Fig.~\ref{#1}}}
\newcommand{\tableref}[1]{Table~\ref{#1}}

\begin{document}

\newcommand{\sys}{\mbox{\textsc{DIP}}}

\date{}



\newcommand{\slogan}{Efficient Large Multimodal Model Training with Dynamic Interleaved Pipeline}

\title
[\sys{}: \slogan]
{\sys{}: \slogan}

\author{Zhenliang Xue}
\orcid{0009-0004-4904-0078}
\affiliation{%
    \department{Institute of Parallel and Distributed Systems}
    \institution{Shanghai Jiao Tong University}
    \city{Shanghai}
    \country{China}}
\email{xuezhenliang@sjtu.edu.cn}

\author{Hanpeng Hu}
\orcid{0009-0008-5787-5226}
\affiliation{%
    \institution{StepFun}
    \city{Shanghai}
    \country{China}}
\email{haaanpeng@outlook.com}

\author{Xing Chen}
\orcid{0000-0002-6453-5026}
\affiliation{%
    \institution{StepFun}
    \city{Shanghai}
    \country{China}}
\email{xchen382@asu.edu}

\author{Yimin Jiang}
\orcid{0009-0001-0049-873X}
\affiliation{%
    \institution{StepFun}
    \city{Shanghai}
    \country{China}}
\email{jymthu@gmail.com}

\author{Yixin Song}
\orcid{0009-0001-4605-7382}
\affiliation{%
    \department{Institute of Parallel and Distributed Systems}
    \institution{Shanghai Jiao Tong University}
    \city{Shanghai}
    \country{China}}
\email{yixinsong@sjtu.edu.cn}

\author{Zeyu Mi}
\orcid{0000-0001-8395-1319}
\affiliation{%
    \department{Institute of Parallel and Distributed Systems}
    \institution{Shanghai Jiao Tong University}
    \city{Shanghai}
    \country{China}}
\email{yzmizeyu@sjtu.edu.cn}

\author{Yibo Zhu}
\orcid{0000-0002-9113-2660}
\affiliation{%
    \institution{StepFun}
    \city{Shanghai}
    \country{China}}
\email{zhuyibo@stepfun.com}

\author{Daxin Jiang}
\orcid{0000-0002-6657-5806}
\affiliation{%
    \institution{StepFun}
    \city{Shanghai}
    \country{China}}
\email{djiang@stepfun.com}

\author{Yubin Xia}
\orcid{0000-0001-6558-5298}
\affiliation{%
    \department{Institute of Parallel and Distributed Systems}
    \institution{Shanghai Jiao Tong University}
    \city{Shanghai}
    \country{China}}
\email{xiayubin@sjtu.edu.cn}

\author{Haibo Chen}
\orcid{0000-0002-9720-0361}
\affiliation{%
    \department{Institute of Parallel and Distributed Systems}
    \institution{Shanghai Jiao Tong University}
    \city{Shanghai}
    \country{China}}
\email{haibochen@sjtu.edu.cn}

\renewcommand{\shortauthors}{Zhenliang Xue et al.}





\begin{abstract}

Large multimodal models (LMMs) have demonstrated excellent capabilities in both understanding and generation tasks with various modalities. While these models can accept flexible combinations of input data, their training efficiency suffers from two major issues: pipeline stage imbalance caused by heterogeneous model architectures, and training data dynamicity stemming from the diversity of multimodal data.

In this paper, we present \sys{}, a dynamic and modality-aware pipeline scheduling framework designed for LMM training. \sys{} tackles the challenge of \emph{dynamic imbalance} via two key techniques:
(1) separating computations of different modalities into dedicated \emph{pipeline segments} to balance workloads within a continuous set of stages;
(2) dynamically splitting input data into finer-grained, modality-specific \emph{sub-microbatches} to balance workloads across these segments.
By asynchronously generating pipeline schedules on idle CPU resources during training, \sys{} dynamically tailors stage executions to each input batch without stalling the training process.
We validate \sys{} on a diverse set of five LMMs, ranging from 12B to 94B parameters and including vision-language and diffusion models.
Experimental results show that our system achieves up to 97.3\% higher throughput compared to state-of-the-art systems,
demonstrating strong adaptability to fluctuating multimodal training workloads.

\end{abstract}

\maketitle

\section{Introduction}

Transformer-based models~\cite{transformer-nips17,qwen2-5-arxiv25,deepseek-v3-arxiv25,llama3-arxiv24} have demonstrated remarkable capabilities in multimodal understanding, reasoning, and generation, establishing themselves as foundational architectures for next-generation large multimodal models (LMMs)~\cite{qwen2.5-vl-arxiv25,gpt4o-arxiv24,llama3-arxiv24,step-audio-arxiv25,step-t2v-arxiv25,gemini-3-2025}. Modern LMMs integrate multiple \emph{modality modules}, including encoders, decoders, and backbone models connected via modality adapters. This architectural design enables flexible and seamless processing of interleaved multimodal data (\eg text, images, video), thereby supporting diverse task paradigms such as multimodal document understanding~\cite{mlongdoc-arxiv24} and multi-turn dialogs~\cite{gpt4o-image-generation}.

However, this architectural flexibility introduces significant challenges in training due to irregular and fluctuating execution latencies across different modality modules and data batches, leading to a unique problem of \emph{dynamic imbalance}. This challenge stems from two interrelated factors:

\textbf{Pipeline Stage Imbalance:}
LMMs exhibit architectural heterogeneity arising from diverse modality modules with distinct parameter shapes and operator types (\eg attention, GEMM, convolution). These differences create skewed computational workloads and varying memory access patterns, complicating the partitioning of model execution into balanced pipeline stages.
Even with exhaustive enumeration of all possible layer splits, the ``optimal'' partitioning still incurs a 22.8\% pipeline bubble overhead for a 37B-parameter LMM (\S\ref{section:impact-on-pipeline-parallelism}).
This inefficiency is primarily due to the significant discrepancy between heterogeneous layers, which prevents perfectly balanced pipeline stage partitioning.

\textbf{Training Data Dynamicity:}
The inherent architectural heterogeneity is further exacerbated by the diversity of multimodal training data, which comprises large-scale datasets with highly variable modality distributions across batches.
Although data packing techniques~\cite{wlb-llm-osdi25,dynapipe-eurosys24} aim to produce more balanced batches, computational imbalance persists. In our experiments, the largest data sample incurs $4.15\times$ greater computational load than the smallest (\S\ref{section:dynamic-imbalance}). Such disparities cause substantial workload fluctuations between batches.

The combination of pipeline stage imbalance and training data dynamicity results in a significant performance bottleneck. Our experiments show that this dynamic imbalance can inflate training overhead by up to 40.3\% for the 37B LMM (\S\ref{section:impact-on-pipeline-parallelism}).
This performance degradation renders the conventional approach of using a single, static pipeline schedule (\eg Megatron-LM's 1F1B) impractical and inefficient for the dynamic nature of LMMs.
Similarly, training systems for multimodal models such as Spindle~\cite{spindle-asplos25} and Optimus~\cite{optimus-atc25} rely on static pipeline schedules tailored for a fixed set of tasks.
Their training plans are determined before training start, overlooking the dynamic nature of multimodal data and resulting in suboptimal pipeline performance.
Existing approaches~\cite{wlb-llm-osdi25,dynapipe-eurosys24,flexpipe-atc25} designed for dynamic text sequence lengths are also insufficient.
They primarily address workload variations that affect all layers of a unimodal LLM uniformly.
However, they cannot resolve the fundamental inter-modality imbalance in LMMs,
where a change in one modality (\eg more images) can drastically overload a specific module while leaving others underutilized.

To address these challenges, we propose \sys{}, a dynamic and modality-aware scheduling framework designed to optimize pipeline parallelism for LMMs. \sys{} is built upon two key insights that identify the root causes of pipeline inefficiency.
First, co-locating computations from different modality modules within the same pipeline execution pass is inherently inefficient due to their disparate computational costs (\figref{figure:schedule-observations}a).
We define a \emph{pipeline segment} as a complete forward or backward pass of computations across all pipeline stages.
To eliminate this \emph{intra-segment imbalance}, \sys{} enforces separated partitioning, dedicating distinct pipeline segments to each modality (\figref{figure:schedule-observations}b).
This isolation prevents slow modalities from bottlenecking faster ones.
Second, even after separation, workload imbalances persist between these modality-specific segments.
To resolve this \emph{inter-segment imbalance}, \sys{} further dynamically splits data into smaller, modality-specific \emph{sub-microbatches} (\figref{figure:schedule-observations}c).
This modality-aware partitioning approach allows latencies of different modality segments to be more closely matched,
resulting in finer-grained load balancing between these segments.

\begin{figure}[!t]
    \includegraphics[width=1\columnwidth]{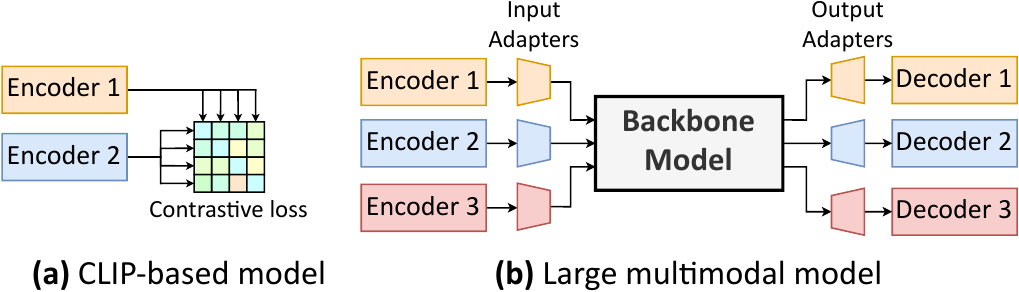}
    \caption{
        Comparison between CLIP-based models and LMMs.
    }
    \label{figure:mm-vs-lmm}
\end{figure}

\begin{figure}[!t]
    \includegraphics[width=1\columnwidth]{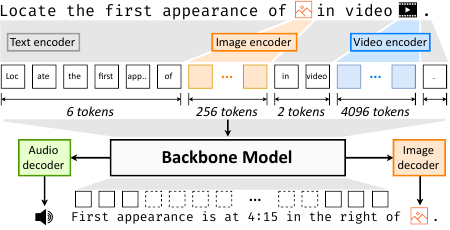}
    \caption{
        An example of LMM that uses a large language model as the backbone.
        The user prompt consists of an image and a video clip,
        and is converted into tokens by corresponding modality encoders,
        which are further processed by the backbone model to produce the response
        in multimodal text or speech audio.
    }
    \label{figure:mllm-demo}
\end{figure}

Meanwhile, in contrast to conventional systems that rely on static or pre-computed schedules, \sys{} employs asynchronous schedule generation. It generates a custom pipeline schedule for each data batch on-the-fly, seamlessly adapting to the fluctuating computational demands of multimodal data.
To avoid stalling training, \sys{} generates schedules using idle CPU cores in parallel with the main GPU training workers, effectively hiding search latency.
To identify high-quality schedules within the tight time constraints of a training iteration (typically 10--60 seconds), \sys{} decomposes the complex search problem into three simpler subproblems,
each with a tailored heuristic to reduce search space.
\sys{} parallelizes the search loop over hundreds of CPU cores, shortening the search time and harnessing idle CPUs.

We implement \sys{} on top of Megatron-LM, the state-of-the-art distributed training framework for transformer-based large language models. In addition to the scheduling algorithms, we develop a training simulator for fast and accurate prediction of pipeline stage latency and memory consumption. We also enhance Megatron-LM with a reconfigurable pipeline mechanism to support dynamic pipeline schedule deployment.
We validate \sys{} on a diverse set of five LMMs, ranging from 12B to 94B parameters and including vision-language models and diffusion models.
Experimental results demonstrate that \sys{} achieves up to 97.3\% higher training performance compared to baseline systems.
Moreover, \sys{} exhibits excellent adaptability to the dynamic workloads of LMM training, maintaining near-optimal hardware utilization throughout the training process.

In summary, this paper makes the following contributions:

\begin{itemize}
    \item We identify and characterize the problem of dynamic imbalance in LMM training, a challenge arising from the interplay between pipeline stage imbalance and training data dynamicity.
    \item We propose \sys{}, a novel scheduling framework that employs asynchronous schedule generation, modality-aware partitioning, and a decomposed search algorithm to dynamically adapt to varying workloads.
    \item We evaluate \sys{} against three state-of-the-art training systems across five LMM models up to 94B, demonstrating improvements of up to 97.3\% in training efficiency.
\end{itemize}

\section{Dynamic Imbalance Characterization}
\label{section:background}

\begin{figure}[t]
    \includegraphics[width=1\columnwidth]{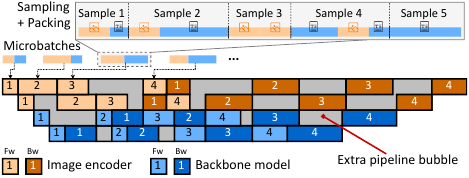}
    \caption{
        Illustration of the impact of dynamic imbalance.
    }
    \label{figure:imbalanced-pipeline}
\end{figure}

\begin{figure*}[!t]
    \includegraphics[width=1\textwidth]{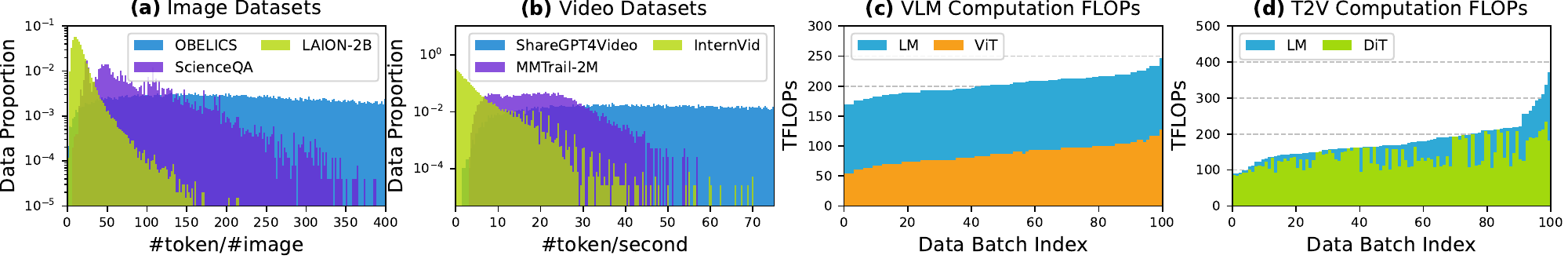}
    \caption{
        \textbf{(a--b)} Token distribution per image in OBELICS~\cite{obelics-arxiv23}, LAION-2B~\cite{laion5b-nips22}, and ScienceQA~\cite{scienceqa-nips22} datasets, and token distribution per second in ShareGPT4Video~\cite{share-gpt4-video-arxiv24}, InternVid~\cite{internvid-arxiv23}, and MMTrail-2M~\cite{mmtrail-arxiv24} video datasets. The Y-axis shows normalized data proportions.\linebreak
        \textbf{(c--d)} Computational requirements for VLM 12B (VLM-S in Table~\ref{table:model-setup}) and T2V 13B (T2V-S in Table~\ref{table:model-setup}) models across 100 packed data batches. Batches (X-axis) are sorted by ascending computational cost, with floating-point operations (Y-axis) measured in teraFLOPs (TFLOPs).
    }
    \label{figure:data-distribution}
\end{figure*}

\subsection{Large Multimodal Models}

Multimodal models have long sought to bridge heterogeneous modalities
by learning unified semantic representations~\cite{modality-bridge-arxiv23}.
Early CLIP-based multimodal models~\cite{clip-arxiv21} pioneered the dual-encoder architecture (\eg ViT for images, BERT for text) and employed contrastive loss~\cite{simclr-icml20} to align cross-modal embeddings in shared semantic spaces.

The advent of large multimodal models (LMM) marks a paradigm shift that extends beyond representation alignment to encompass multimodal generation capabilities. This evolution is driven by the integration of large-scale transformer-based architectures, such as large language models (LLMs), vision transformers (ViT), and diffusion transformers (DiT).
LMMs integrate encoders with the backbone model using modality adapters, as illustrated in \figref{figure:mm-vs-lmm}. The backbone model subsequently generates output representations either through autoregressive processes or diffusion mechanisms. These representations are then converted into human-perceivable formats, including text, audio, and images, via specialized modality decoders.
The scalable and generative architecture of LMM significantly broadens the spectrum of supported tasks compared to CLIP-based models.
For instance, in an LMM (\figref{figure:mllm-demo}), users can interleave text, images, and video within a single query and iteratively refine their queries by appending follow-up questions to create multi-turn dialogs~\cite{gpt4o-image-generation,gemini-3-2025,dialog-gen-arxiv24}.
The LMM can respond in text or speech~\cite{step-audio-arxiv25}, and generate images using diffusion modules~\cite{moviegen-arxiv24,step-t2v-arxiv25,hunyuan-dit-arxiv24}.

\heading{LMM Training}
Large transformer-based models are typically trained with 3D distributed schemes such as data parallelism (DP), pipeline parallelism (PP), and tensor parallelism (TP)~\cite{megatron-lm-arxiv20}.
Pipeline parallelism partitions model layers into multiple \emph{model chunks} placed on different machines (\ie pipeline ranks).
Each model chunk can perform both forward and backward stage computations.
Data batches are split into smaller microbatches and passed between pipeline stages with point-to-point (P2P) communications.

\subsection{Sources of Dynamic Imbalance}
\label{section:dynamic-imbalance}

Due to the heterogeneity of multimodal model architectures and the diversity of training data,
the training process for LMMs differs significantly from that for unimodal models in
\emph{pipeline stage imbalance} and \emph{training data dynamicity}.

\heading{Pipeline Stage Imbalance}
Optimal pipeline efficiency requires balanced stage partitioning to minimize bubbles, yet inherent computational disparities between modality modules introduce fundamental challenges.
Consider a 37B VLM comprising a 5B ViT encoder (64 layers) and 32B language model (64 layers) on H800 GPUs,
processing a batch with 8 images and 8192 text tokens.
Each ViT layer processes the images in 6.75ms (forward+backward), while each LM layer requires 10.5ms for the tokens.
The optimal partitioning scheme produced by exhaustive search across 16 pipeline stages yields stage latencies between 63ms and 73.5ms (\ie 16.7\% variation).
With 64 microbatches and Megatron-LM's 1F1B pipeline schedule, this imbalance introduces 22.8\% additional pipeline bubbles, demonstrating the inherent difficulty in achieving perfectly balanced partitioning.

\heading{Training Data Dynamicity}
This challenge is further compounded by the dynamic heterogeneity in multimodal training data,
where the variability across batches from different modalities induces fluctuating computational demands.
Multimodal data originates from diverse sources including:
image-description pairs~\cite{laion5b-nips22,coyo-700m},
video with captions~\cite{share-gpt4-video-arxiv24,internvid-arxiv23,mmtrail-arxiv24,msr-vtt-cvpr16}, and
interleaved image-text content~\cite{obelics-arxiv23,mmc4-arxiv23}.
Data samples typically comprise images/videos with descriptive texts, exhibiting significant \emph{cross-dataset variation} in modality data ratios (\figref{figure:data-distribution}a--b).
For instance,
the LAION-2B~\cite{laion5b-nips22} dataset consists of images paired with short captions, with a small text-image ratio (16.4 tokens/image), while the OBELICS~\cite{obelics-arxiv23} dataset contains full-length multimodal documents
with highly variable ratios (0.4 to 3115 tokens/image).

This variation further induces \emph{cross-batch imbalance} during data packing.
Typically, data samples are packed into larger batches to enhance computation efficiency~\cite{dynapipe-eurosys24,wlb-llm-osdi25}.
In vision-language models (VLMs),
both images and texts are tokenized (\eg one image into 169 patch tokens)
and then greedily packed up to the model's context length (\eg 8192 tokens) to form a microbatch.
Similarly, for text-to-video (T2V) diffusion models,
video clips with similar durations and aspect ratios are often grouped for batch processing~\cite{moviegen-arxiv24,step-t2v-arxiv25}.
However, data packing fails to ensure balanced workloads across multiple modalities
due to their divergent distributions (\figref{figure:data-distribution}c--d).
For instance, consider two microbatches with same 8192-token capacity:
the first contains 10 images (\ie 1690 patch tokens) accompanied by 6502 text tokens,
whereas the second consists of only a single image and 8023 text tokens.
These two microbatches impose disparate computational demands:
the FLOPs required for the image pipeline stages in the first microbatch are $10\times$ higher than those in the second.
This discrepancy leads to imbalanced pipeline stages across microbatches (\figref{figure:imbalanced-pipeline}).
Furthermore, for a T2V model comprising a 7B LM and a 5B DiT,
the most computationally intensive batch (data batch index 100 in \figref{figure:data-distribution}d) imposes a $4.15\times$ greater computational load than the smallest batch (data batch index 1),
even after data packing.


\subsection{Negative Impact on Pipeline Parallelism}
\label{section:impact-on-pipeline-parallelism}

The combination of pipeline stage imbalance and training data dynamicity leads to the \emph{dynamic imbalance} problem.
As depicted in \figref{figure:imbalanced-pipeline}, this issue introduces significant pipeline bubbles in Megatron-LM's 1F1B pipeline schedule, and severely degrades training throughput.

To quantify this impact, we compare two model setups with the same number of parameters:
a unimodal LM (7B) and
a vision-language model (ViT 2B + LM 5B).
Under identical 1F1B pipeline configurations
(using balanced parameter partitioning and fixed computational budgets),
the VLM incurs a 12.5\% overhead on static data due to stage imbalance.
With real-world dynamic data, this overhead escalates to 40.3\%,
as shown in \tableref{table:imbalanced-pipeline}.

\begin{table}[t]
    \caption{
        Training performance of 7B models on 8 GPUs (TP=2, PP=4).
        ``PFLOPs'' denotes the floating-point operations per iteration in petaFLOPs,
        which is controlled to a fixed value for fair comparison.
        ``MFU'' refers to model FLOPs utilization.
    }
    \label{table:imbalanced-pipeline}
    \small
    \begin{tabular}{>{\raggedright}p{4cm}ccc}
    \toprule
    \textbf{Model Setup} & \textbf{Time (s)} & \textbf{PFLOPs} & \textbf{MFU} \\
    \midrule
    LM 7B & 4.068 & 12.8 & 0.400 \\
    ViT 2B + LM 5B (static data) & 4.567 & 12.7 & 0.351 \\
    ViT 2B + LM 5B (dynamic data) & 6.789 & 12.8 & 0.239 \\
    \bottomrule
    \end{tabular}
\end{table}

\heading{Impact on Pipeline Design}
Such performance degradation renders the fixed pipeline schedules used in unimodal LLMs impractical.
Since each iteration processes a distinct data batch, the optimal pipeline schedule changes dynamically.

Precomputation of pipeline schedules is also infeasible. The number of possible input combinations grows exponentially in two dimensions: the number of modalities and the number of microbatches. In our experimental setup, each microbatch can contain up to 48 images (\S\ref{section:datasets}), leading to $49^{64} \approx 1.5\mathrm{e}108$ distinct input configurations for 64 microbatches. This makes it impossible to exhaustively precompute optimal schedules for all scenarios. Although precomputing over a subset of inputs could reduce overhead, selecting a representative subset from this enormous space is highly challenging, and fails to generalize to unseen inputs.

Several prior works have addressed the issue of dynamicity in text sequence lengths,
by improving data packing strategies to reduce discrepancies between microbatches~\cite{wlb-llm-osdi25},
or adaptively reordering microbatches to minimize bubbles caused by irregular inputs~\cite{dynapipe-eurosys24,flexpipe-atc25}.
However, these methods are inadequate for solving dynamic imbalance in multimodal training, as they overlook the heterogeneity among modality modules.
Unlike in unimodal LLMs, where sequence length variations affect all layers uniformly, multimodal microbatches affect modality-specific modules differently. For example, increasing the number of images substantially raises computational demands on image encoders, while barely affecting the language model backbone. This inter-modality imbalance makes data-centric approaches designed for unimodal models ineffective for LMM training.

\section{Approach Overview}
\label{section:overview}

To address the challenges of pipeline stage imbalance and training data dynamicity,
we introduce \underline{D}ynamic \underline{I}nterleaved \underline{P}ipeline (\sys{}), a dynamic and modality-aware scheduling framework designed to optimize pipeline parallelism for LMMs.
\sys{} is built on three key design principles:
asynchronous schedule generation to hide search latency,
modality-aware partitioning to mitigate imbalance at its source, and
decomposed search algorithm to find high-quality pipeline schedules efficiently.

\subsection{Design Principles}
\label{section:design-decisions}

\begin{figure}[b]
    \includegraphics[width=1.0\columnwidth]{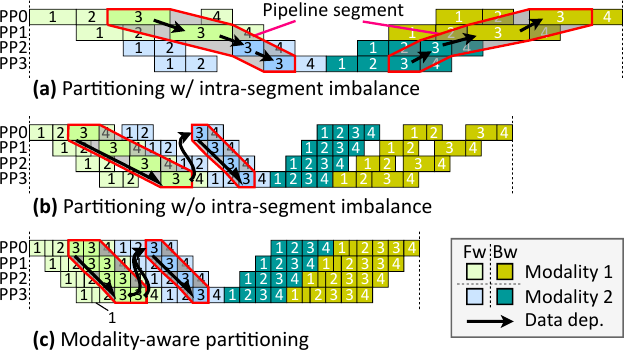}
    \caption{
        Illustrations of partitioning schemes from \S\ref{section:design-decisions}.
        Numbers in boxes denote microbatches.
        \textbf{(a)} Non-modality-aware partitioning co-locates stages from different modalities in the same pipeline segment, causing intra-segment imbalance.
        \textbf{(b)} Separated partitioning dedicates distinct pipeline segments to each modality, removing intra-segment imbalance.
        \textbf{(c)} Modality-aware data batching further splits microbatches into smaller, modality-specific sub-microbatches to balance workloads across segments.
    }
    \label{figure:schedule-observations}
\end{figure}

\heading{Asynchronous Schedule Generation}
In contrast to conventional approach\-es that rely on static or pre-computed schedules, \sys{} adopts an online, asynchronous strategy,
in order to handle the dynamic nature of LMM data.
For each upcoming data batch, it concurrently generates a tailored pipeline schedule on-the-fly, effectively adapting to the fluctuating computational demands of multimodal data.
This process, which includes data prefetching and schedule searching, runs asynchronously on idle CPU resources, parallel to the main training workers on GPUs. By staying off the critical path, our approach provides the adaptability needed for dynamic data without introducing significant overhead.

\heading{Modality-Aware Partitioning}
The primary source of inefficiency in LMM training stems from the dynamic computational imbalance between different modality modules, as discussed in \S\ref{section:dynamic-imbalance}.
Such dynamic imbalance creates irregular stage latencies and makes it difficult to remove pipeline bubbles.
To address this, \sys{} aims to reduce pipeline bubbles at their source by minimizing stage latency discrepancies.
This is achieved through partitioning the computation of different modality modules into equally-wide \emph{pipeline segments}.
Our approach is guided by two key techniques:

\textbf{\ding{172} Remove Intra-Segment Imbalance with Separated Partitioning.}
A \emph{pipeline segment} is a set of consecutive stages distributed across all $P$ pipeline ranks.
For example, the four forward stages numbered with ``3'' in \figref{figure:schedule-observations}a form one pipeline segment,
while their corresponding backward stages form another one.
We observe that co-locating stages from different modality modules within the same segment (\figref{figure:schedule-observations}a) is inherently inefficient. Because vision and language stages have different computational costs and sensitivities to data proportions (\eg number of images vs. tokens), mixing them creates unavoidable latency disparities between pipeline ranks. Our first principle is to enforce \emph{separated partitioning} (\figref{figure:schedule-observations}b), where each modality module occupies its own dedicated pipeline segments. This eliminates a primary source of pipeline bubbles. For the ViT 2B + LM 5B model (\S\ref{section:dynamic-imbalance}), this strategy alone improves performance by 13.1\% over the best mixed partitioning scheme.

\textbf{\ding{173} Remove Inter-Segment Imbalance with Modality-Aware Batching.}
After separating modalities, we must balance the workload between their respective segments. Our second principle is to partition data within a microbatch into smaller, modality-specific \emph{sub-microbatches} (\figref{figure:schedule-observations}c). For example, if the entire vision encoder stage is slower than a LM stage, we can process images in smaller sub-microbatches.
This allows the encoder to execute multiple shorter stages, and each encoder stage latency can be tailored to better match the latency of a single LM stage.
This modality-aware batching reduces latency discrepancies across the entire pipeline, enabling a more globally balanced and efficient schedule.

\begin{figure}[t]
    \includegraphics[width=1\columnwidth]{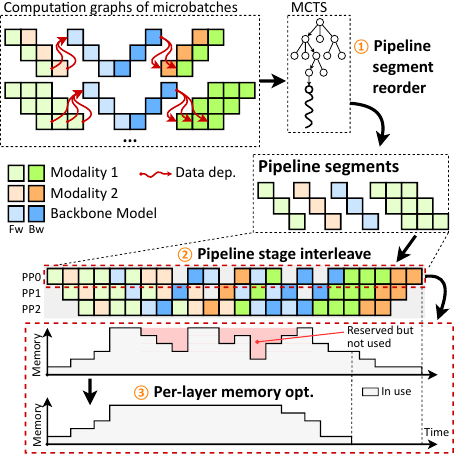}
    \caption{
        The workflow of \sys{}'s pipeline schedule searcher.
    }
    \label{figure:plan-searcher}
\end{figure}

\heading{Decomposed and Scalable Schedule Search}
The scheduling search problem can be formulated as a monolithic integer linear program (ILP) and solved with off-the-shelf solvers~\cite{highs,gurobi}.
This approach is precise but intractable for real-time use, often taking minutes or hours to solve~\cite{recycle-sosp24}.
To meet the tight time budget of a single training iteration (typically 10--60 seconds, \figref{figure:e2e-perf}b), we adopt a divide-and-conquer strategy that decomposes the complex search problem into a sequence of simpler, more manageable subproblems. As shown in \figref{figure:plan-searcher}, our searcher iterates through a three-phase loop, applying tailored heuristics at each step:

\textbf{\ding{172} Pipeline Segment Reordering (\S\ref{section:segment-reordering}):} First, we determine the optimal processing sequence of pipeline segments. We use Monte Carlo Tree Search (MCTS) to efficiently explore the vast permutation space and identify promising pipeline segment orderings.

\textbf{\ding{173} Pipeline Stage Interleaving (\S\ref{section:stage-interleaving}):} With a fixed segment order, we then arrange the corresponding forward and backward pipeline stages. A fast, dual-queue greedy algorithm interleaves these stages to minimize pipeline bubbles and pack the schedule tightly.

\textbf{\ding{174} Per-Layer Memory Optimization (\S\ref{section:memory-optimization}):} Finally, with the schedule structure fixed, we optimize memory usage independently on each pipeline rank. This phase selects memory-saving strategies (e.g., activation checkpointing and offloading) for all model layers to minimize memory usage fluctuations induced by data dynamicity and optimize end-to-end iteration latencies.

This entire loop is highly parallelizable across CPU threads, ensuring that our search process scales with available CPU resources and can serve large distributed training clusters.

\subsection{System Workflow}
\label{section:dip-overview}

The principles above are orchestrated by the \sys{} training planner, which consists of three main components: modality-aware partitioner (\S\ref{section:modality-aware-partitioning}), pipeline schedule searcher (\S\ref{section:pipeline-schedule-searcher}), and training simulator (\S\ref{section:multimodal-training-simulator}).

\begin{figure}[t]
    \includegraphics[width=0.95\columnwidth]{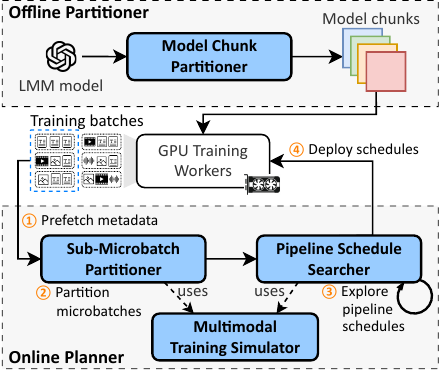}
    \caption{
        The overall workflow of \sys{}.
        Before training, the LMM model is partitioned into model chunks by the model chunk partitioner.
        During each training iteration, \sys{} asynchronously executes a four-stage online planning process.
    }
    \label{figure:overview}
\end{figure}

The overall workflow proceeds in two phases, as shown in \figref{figure:overview}.
The modality-aware partitioner comprises an offline model chunk partitioner and an online sub-microbatch partitioner.
\textbf{Before training,} the model chunk partitioner divides the LMM into model chunks based on our separated partitioning principle and distributes them to the designated GPUs.
\textbf{During training,} all model chunks remain statically placed on their respective GPU workers, which also fixes the underlying network topology (\eg NCCL connections).
For each iteration, the planner executes a four-stage process:

\textbf{\ding{172} Metadata Prefetching:} The planner fetches metadata for the next batch (\eg token counts, number of images).

\textbf{\ding{173} Adaptive Microbatch Partitioning:} Using this metadata, the sub-microbatch partitioner divides the microbatches into smaller, modality-specific sub-microbatches to enable inter-segment load balancing.

\textbf{\ding{174} Pipeline Schedule Search:} Pipeline schedule searcher is guided by performance estimates from the simulator and runs its three-phase algorithm to find an optimal schedule for the generated sub-microbatches.

\textbf{\ding{175} Runtime Deployment:} The best schedule is compiled into an execution plan and dispatched to the distributed training runtime for execution.


\section{Modality-Aware Partitioner}
\label{section:modality-aware-partitioning}

Building on the principles in \S\ref{section:design-decisions},
we propose \emph{modality-aware partitioning}.
This approach partitions each modality module into multiple model chunks before training,
and dynamically splits microbatches into modality-specific sub-microbatches during training.
Specifically, for the $i$-th modality, {\sys} determines two parameters:
the sub-microbatch size $B_i$ and the number of pipeline segments $K_i$.
Consequently, the $i$-th modality module is partitioned into $P \cdot K_i$ model chunks,
distributed across $P$ pipeline ranks.

\heading{Determine Sub-Microbatch Size}
\sys{} independently selects the minimum viable sub-microbatch size $B_i$ for each modality module through systematic profiling.
Although smaller sub-microbatch sizes enable finer-grained pipeline partitioning and higher scheduling efficiency,
excessively small sizes may lead to GPU underutilization (\figref{figure:pipeline-scheduling}).
To balance these trade-offs, \sys{} measures the computation latency of each modality module across varying sub-microbatch sizes.
The optimal $B_i$ is determined as the smallest sub-microbatch size that maintains at least 95\%
of the peak GPU computational efficiency observed across all tested configurations.

\heading{Partition Model Chunks}
After determining sub-microbatch sizes, \sys{} partitions each modality module to achieve global latency balancing.
Given $n$ modality modules with sorted computation latencies $T_1 \leq T_2 \leq \cdots \leq T_n$
(measured at their respective $B_i$ sizes),
\sys{} assigns pipeline segment counts proportional to these latencies.
Specifically, the fastest module ($T_1$) is configured as a single pipeline segment,
while the $i$-th module is partitioned into $K_i = \lfloor T_i / T_1 \rfloor$ segments.
For a modality module comprising $L_i$ layers,
\sys{} distributes layers across $P \cdot K_i$ model chunks,
with each chunk containing $L_i/(P \cdot K_i)$ consecutive layers.

\heading{Construct Sub-Microbatch}
During training execution, \sys{} dynamically constructs sub-microbatches for each modality module
based on the content of incoming microbatches.
For a microbatch assigned to the $i$-th modality module with $N_i$ instances,
\sys{} splits it into $M_i = \lceil N_i/B_i \rceil$ uniformly partitioned sub-microbatches.
This process generates $2 M_i \cdot K_i$ pipeline segments in total
(accounting for both forward and backward computations) for the modality module.

\section{Pipeline Schedule Searcher}
\label{section:pipeline-schedule-searcher}

\subsection{MCTS-Based Pipeline Segment Reordering}
\label{section:segment-reordering}

\sys{} determines the relative ordering between pipeline segments by assigning scheduling priorities to them.
These priorities are subsequently utilized in the pipeline stage interleaving phase (\S\ref{section:stage-interleaving}).
To efficiently explore the search space of possible priority assignments, \sys{} employs Monte Carlo tree search (MCTS)~\cite{mcts}.

\heading{Search Tree Construction}
Given $n$ pipeline segments, MCTS constructs a sequence of length $n$ by iteratively selecting the segment for each position $i$ (from $1$ to $n$).
The segment at the $i$-th position receives a priority of $n - i$, ensuring earlier selected segments obtain higher priorities.

The algorithm builds a search tree where each node at depth $d$ corresponds to the $d$-th element in the sequence.
Consequently, any root-to-node path ending at depth $d$ represents a partial sequence for the first $d$ positions.
Each node $v$ maintains the highest performance score $s_v$ observed among its descendants in the MCTS tree.

\heading{Search Loop}
MCTS iteratively performs the following four phases until convergence:

\begin{enumerate}
    \item \textbf{Node Selection:} Starting from the root, MCTS navigates downward by selecting child nodes that maximize the upper confidence bound (UCB) score $s_v^\alpha + \beta\sqrt{(\log{N_x}) / N_v}$.
    Here, $x$ denotes the current node, $v$ is the target child node, $N_x$ and $N_v$ are visit counts, and $\alpha$, $\beta$ are hyperparameters.
    This process repeats until a leaf node $u$ is reached.

    \item \textbf{Tree Expansion:} A new node representing the next element after $u$ is added to the search tree.
    Then $u$ is set to the newly created node.

    \item \textbf{Random Rollouts:} Multiple rollouts (\eg 10 trials) generate complete sequences by randomly assigning segments to remaining positions after $u$.
    Each sequence then undergoes pipeline stage interleaving (\S\ref{section:stage-interleaving}) and per-layer memory optimization (\S\ref{section:memory-optimization}) to compute performance scores (\ie end-to-end iteration time).

    \item \textbf{Score Backpropagation:} The best rollout score among all trials propagates upward from $u$ to update the performance scores of ancestor nodes.
\end{enumerate}

This loop continues until either a predefined time budget is exhausted or the entire search space is explored.

\heading{Optimization}
We observe that segments processing the same modality within a single microbatch have identical pipeline structures and similar stage latencies.
Consequently, their relative ordering does not impact end-to-end performance.
This insight allows \sys{} to assign identical priorities to such segments and enforce a fixed ordering between them, significantly reducing the search space.

\subsection{Greedy Pipeline Stage Interleaving}
\label{section:stage-interleaving}

After assigning priorities to segments,
\sys{} employs a dual-queue algorithm to adaptively interleave forward and backward pipeline stages.
When both schedulable forward and backward stages are available,
\sys{} mimics Megatron-LM's memory-efficient ``one-\-forward-\-one-\-backward'' (1F1B) pattern~\cite{megatron-lm-arxiv20,dynapipe-eurosys24}.
When either type is unavailable,
a greedy strategy fills pipeline bubbles to construct compact schedules.

\heading{Initialization}
\sys{} first obtains the stage latencies and memory consumptions for all stages via the training simulator,
using the most memory-efficient scheme determined in \S\ref{section:memory-optimization}.
This step ensures sufficient optimization space for subsequent per-layer memory optimizations.

For each pipeline rank, \sys{} maintains:
(1) the end time of the last scheduled stage ($t_\mathrm{last}$, initially zero);
(2) two priority queues ($Q_\mathrm{fw}$, $Q_\mathrm{bw}$) storing forward and backward stages in descending priority order; and
(3) the minimum start time ($t_\mathrm{min}$) among stages in $Q_\mathrm{fw}$ and $Q_\mathrm{bw}$.
Each stage maintains a minimum schedulable start time ($t_\mathrm{start}$) initialized to zero for stages with no predecessors and $+\infty$ otherwise.

\heading{Iterative Scheduling}
\sys{} iteratively selects a pipeline rank and schedules one stage from it:
\begin{enumerate}
    \item Select the pipeline rank with the smallest $t_\mathrm{min}$.
    \item Compare the minimum start times of stages in $Q_\mathrm{fw}$ ($t_\mathrm{fw}$) and $Q_\mathrm{bw}$ ($t_\mathrm{bw}$) against $t_\mathrm{last}$.
    \item If both $t_\mathrm{fw} < t_\mathrm{last}$ and $t_\mathrm{bw} < t_\mathrm{last}$,
          schedule alternating forward/backward stages based on the last scheduled stage's computation type,
          emulating the 1F1B pattern.
    \item Otherwise, select the stage with the smallest $t_\mathrm{start}$ to minimize pipeline bubble between $t_\mathrm{min}$ and stage's start time.
\end{enumerate}

The selected stage is dequeued and scheduled on the pipeline rank.
Subsequently, $t_\mathrm{last}$, $t_\mathrm{min}$, and $t_\mathrm{start}$ values for all successor stages are updated.

\heading{Memory Constraints}
Throughout scheduling, \sys{} tracks real-time memory consumption across pipeline ranks.
When a rank exceeds memory capacity, its forward queue is temporarily disabled to prevent memory overflow.

\subsection{Per-Layer Memory Optimization}
\label{section:memory-optimization}

\sys{} adaptively selects appropriate memory optimization strategies (\eg activation checkpointing)
for all model layers.
Since the stage interleaving scheme is predetermined by the dual-queue algorithm (\S\ref{section:stage-interleaving}),
this phase independently optimizes end-to-end latency for each pipeline rank.

\heading{Offline Candidate Generation}
Before training, \sys{} enumerates all possible optimization strategies for each model layer
and estimates their execution time and memory consumption using a training simulator.
The system then groups each forward stage with its corresponding backward stage into a \emph{stage pair}.
For each stage pair, \sys{} selects up to $S$ candidate strategies (\eg $S = 10$)
from the combinatorial space of layer-wise strategies through a three-step process:
(1) identifying the fastest candidate,
(2) identifying the most memory-efficient candidate, and
(3) evenly partitioning the memory range between these extremes into $S-2$ buckets,
then selecting the most time-efficient candidate within each bucket via a multiple-choice knapsack algorithm~\cite{multi-choice-knapsack}.

\heading{ILP Formulation}
For each pipeline rank, \sys{} employs an ILP solver to select optimal candidates.
Given $n$ stage pairs, each with $S$ candidate strategies, let $s_i$, $t_i$ denote the start and end timestamps of the $i$-th stage pair,
and $\mathrm{lat}_{i,j}$, $\mathrm{mem}_{i,j}$ denote the stage latency and memory consumption of the $j$-th candidate for the $i$-th stage pair.
The formulation uses:

\begin{itemize}
    \item \textbf{Variables:} $o_{i,j} \in \{0,1\}$ indicating whether the $j$-th candidate is selected for the $i$-th stage pair.
    \item \textbf{Objective:} Minimize total latency $\sum_i^n \sum_j^S o_{i,j} \cdot \mathrm{lat}_{i,j}$.
    \item \textbf{Selection Constraints:} Each stage pair selects exactly one strategy, \ie $\sum_j^S o_{i,j} = 1$ ($\forall 1 \leq i \leq n$).
    \item \textbf{Memory Constraints: } At any time $s_k$, the memory limit $M$ is satisfied,
    \ie $\sum_i^n [s_i \leq s_k \leq t_i] \sum_j^S o_{i,j} \cdot \mathrm{mem}_{i,j} \leq M$ ($\forall 1 \leq k \leq n$),
    where $[\cdot]$ is the Iverson bracket~\cite{iverson-bracket}.
\end{itemize}

\heading{Optimizations}
Although the problem scale has been significantly reduced compared to the monolithic, end-to-end ILP formulation,
solving an ILP instance may still require several seconds.
We introduce two techniques to achieve efficient solving ($<$10 ms per instance on a single CPU core):
(1) warm-starting the ILP solver with a greedy initial solution, and
(2) permitting a small optimality gap (\eg $\leq 5\%$) for early termination,
as closing the final 5\% gap incurs diminishing returns but prohibitive computational costs.

\subsection{Time Complexity Analysis}

In this section, we outline the algorithmic structure of {\sys}'s training planner and
analyze its time complexity.
The planner operates as an iterative search loop that continues until a predefined time budget is exhausted.
Each iteration consists of three stages:
pipeline segment reordering (\S\ref{section:segment-reordering}),
pipeline stage interleaving (\S\ref{section:stage-interleaving}), and
per-layer memory optimization (\S\ref{section:memory-optimization}).
For the complexity analysis of a single search iteration, let
$p$ denote the number of pipeline ranks,
$n$ the number of sub-microbatches, and
$S$ the number of candidate memory-saving strategies.
Additionally, let $\mathrm{ApproxILP}(m, k)$ represent the time complexity of the approximate ILP solver for
an instance with $\mathrm O(m)$ variables and $\mathrm O(k)$ constraints.

First, pipeline segment reordering generates an ordering for all sub-microbatches in $\mathrm O(n)$ time.
Second, the greedy scheduling algorithm for pipeline stage interleaving runs in time proportional to the total number of pipeline stages, \ie $\mathrm O(p \cdot n)$.
Finally, per-layer memory optimization solves a set of small, approximate ILP instances to select the optimal configuration for each pipeline rank.
Each ILP instance requires creating $n \cdot S$ indicator variables ($o_{i,j}$),
along with $n$ selection constraints and $n$ memory constraints.
Consequently, the time complexity for per-layer memory optimization is $\mathrm O(p \cdot \mathrm{ApproxILP}(n \cdot S, n))$.
Combining all these steps, the overall time complexity for a single search iteration is $\mathrm O(p \cdot (n + \mathrm{ApproxILP}(n \cdot S, n)))$.

In contrast, a full ILP baseline approach is inherently inefficient due to the massive scale of variables and constraints involved.
First, it requires solving the entire pipeline globally, rather than decomposing the problem per pipeline rank.
Second, it necessitates $\mathrm O(n^2)$ ordering constraints to prevent overlapping stage executions within each pipeline rank, whereas {\sys} avoids this overhead through efficient explicit scheduling.
Furthermore, in the memory optimization phase, assuming $L$ layers per pipeline stage and $c$ candidate strategies per layer, the search space expands to $c^L$ candidates per stage.
Consequently, the complexity of the baseline ILP approach escalates to $\mathrm O(\mathrm{ApproxILP}(p \cdot n \cdot c^L, p \cdot n^2))$.
We empirically compare the performance of this baseline against {\sys}'s training planner in \S\ref{section:planner-evaluation} and \figref{figure:search-scalability}.

\begin{table}[!t]
    \caption{
        Model specifications used in the evaluation.
    }
    \label{table:model-params}

    \footnotesize
    \begin{tabular}{lccccc}
    \toprule
    \textbf{Name} & \textbf{\makecell{\# of\\Layers}} & \textbf{\makecell{Embed\\Dim}} & \textbf{\makecell{FFN\\Hidden\\Dim}} & \textbf{\makecell{\# of\\ Attn.\\Heads}} & \textbf{\makecell{\# of\\ Attn.\\Groups}} \\
    \midrule
    ViT 5B~\cite{vit22b-arxiv23} & 63 & 1792 & 15360 & 16 & 16 \\
    ViT 22B~\cite{vit22b-arxiv23} & 48 & 6144 & 24576 & 48 & 48 \\
    \midrule
    Llama3 8B~\cite{llama3-arxiv24} & 32 & 4096 & 14336 & 32 & 8 \\
    Qwen2 32B~\cite{qwen2-5-arxiv25} & 64 & 5120 & 27648 & 40 & 8 \\
    Qwen2 72B~\cite{qwen2-5-arxiv25} & 80 & 8192 & 29568 & 64 & 8 \\
    \midrule
    DiT 5B~\cite{moviegen-arxiv24} & 28 & 3584 & 10240 & 28 & 28 \\
    DiT 30B~\cite{moviegen-arxiv24} & 48 & 6144 & 24576 & 48 & 48 \\
    \bottomrule
    \end{tabular}
\end{table}

\begin{table}[!t]
    \caption{
        Model combinations used in the evaluation.
    }
    \label{table:model-setup}

    \small
    \begin{tabular}{llccc}
    \toprule
    \textbf{Name} & \textbf{Model Setup} & \textbf{TP} & \textbf{PP} & \textbf{\#GPU} \\
    \midrule
    VLM-S & ViT 5B + Llama3 8B & 4 & 4 & 16 \\
    VLM-M & ViT 5B + Qwen2 32B & 8 & 4 & 32 \\
    VLM-L & ViT 22B + Qwen2 72B & 8 & 8 & 64 \\
    \midrule
    T2V-S & Llama3 8B + DiT 5B & 4 & 4 & 16 \\
    T2V-L & Qwen2 32B + DiT 30B & 8 & 8 & 64 \\
    \bottomrule
    \end{tabular}
\end{table}

\section{Implementation}
\label{section:implementation}

\sys{} is implemented as an extension to Megatron-LM~\cite{megatron-lm-arxiv20},
featuring a central planner that coordinates with Megatron-LM's runtime.
The planner iteratively prefetches training metadata, performs pipeline schedule searches,
and deploys the optimized schedules to GPU clusters.

\subsection{Training Simulator}
\label{section:multimodal-training-simulator}

\sys{} employs operator-level analytical modeling for performance prediction.
The simulator constructs directed acyclic graphs (DAGs) with two node types:
(1) operator nodes representing low-level GPU operations (\eg matrix multiplication, collective communication), and
(2) tensor nodes corresponding to data buffers (\eg model parameters).
Each node is assigned to a specific device (GPU/CPU/NVLink) and connected with dependency edges.

For latency estimation, each operator node is characterized by:
the number of floating-point operations $N_\mathrm{fop}$,
memory accesses $N_\mathrm{mem}$ in bytes,
and network transfers $N_\mathrm{net}$ in bytes.
Given device capabilities (computational capacity $F$ in FLOPS,
memory bandwidth $B_\mathrm{mem}$ in bytes/s,
and network throughput $B_\mathrm{net}$ in bytes/s), the latency is computed as
$
\max\{
    \alpha_\mathrm{fop}N_\mathrm{fop}/F,
    \alpha_\mathrm{mem}N_\mathrm{mem}/B_\mathrm{mem},
    \alpha_\mathrm{net}N_\mathrm{net}/B_\mathrm{net}
\}
$,
where $\alpha_\mathrm{fop}$, $\alpha_\mathrm{mem}$, and $\alpha_\mathrm{net}$ are efficiency scaling factors.
Users may override this cost model with custom estimation rules.

The simulator populates operator timestamps in topological order,
then determines tensor lifetimes by examining related operator nodes.
These lifetimes enable construction of memory consumption timelines
for peak memory usage estimation across devices.

\subsection{Parallel Schedule Search}

\sys{} executes the training simulator and search algorithms on CPU cores.
It begins by simulating individual microbatch computation graphs using prefetched metadata to obtain pipeline stage latencies.
During schedule search, multiple workers explore the schedule space in parallel while sharing global MCTS search statistics.
After each search round, these workers atomically update the global MCTS statistics via mutex-protected operations.
Contention remains minimal because workers perform multiple rollouts between synchronizations, thereby amortizing the overhead.
To avoid interference with normal training processes, \sys{} constrains the search worker count to at most 50\% of available CPU cores
(\eg 64 cores on a machine with 8 H800 GPUs).

\subsection{Execution Plan Deployment}

Deploying simulated pipeline schedules to GPU clusters requires compiling them into physical execution plans
that specify computation and communication patterns.

\heading{Schedule Compilation}
\sys{} defines action sequences constituting pipeline execution plans, following DynaPipe~\cite{dynapipe-eurosys24}:
\begin{itemize}
    \item Pipeline stages translate to \texttt{fw\_stage}/\texttt{bw\_stage} actions with optimization strategies from \S\ref{section:memory-optimization}.
    \item Point-to-point (P2P) communications use asynchronous kernels (\texttt{isend}, \texttt{irecv}) overlapped with computations.
    \item Synchronization actions (\texttt{wait\_isend}, \texttt{wait\_irecv}) are inserted based on simulated timelines.
\end{itemize}

\heading{Runtime Modifications}
We extend Megatron-LM's \texttt{schedules} module to support dynamic plans.
Each pipeline worker receives an action list via RPC from the central planner
and executes it sequentially.
Consecutive P2P kernels are grouped into a batched operation to enhance efficiency.

\begin{figure*}[!t]
    \subfloat[Average end-to-end performance]{
        \includegraphics[height=29mm]{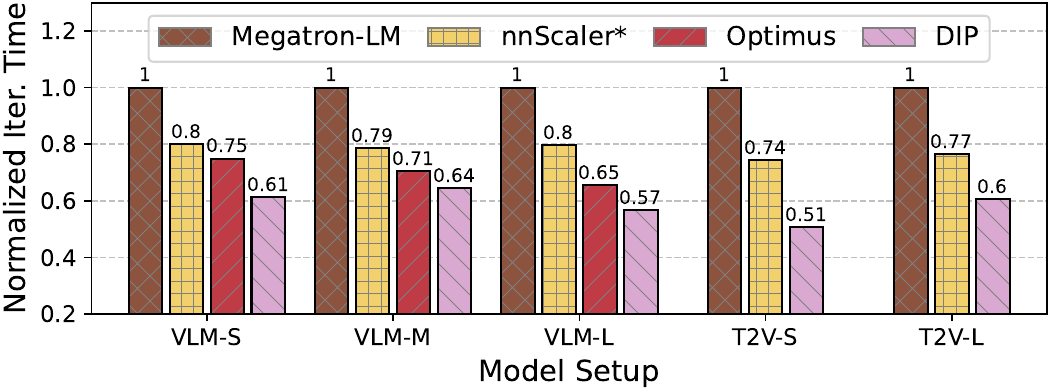}
    }
    \hfil
    \subfloat[Dynamic workloads]{
        \includegraphics[height=29mm]{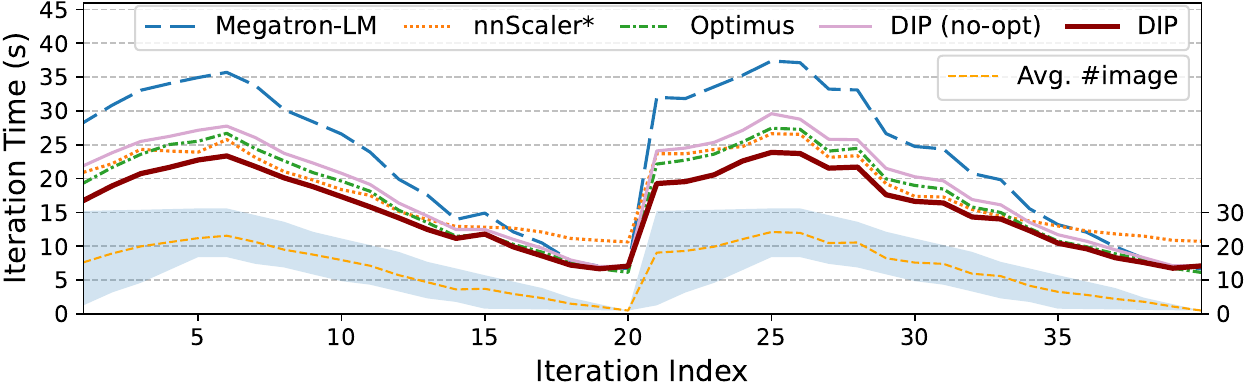}
    }

    \caption{
        End-to-end performance.
        \textbf{(a)} Average performance of 100 iteration on real datasets across five model setups.
        \textbf{(b)} End-to-end latency timeline of 40 consecutive iterations.
        The orange dashed line depicts the average number of images.
    }
    \label{figure:e2e-perf}
\end{figure*}

\section{Evaluation}

\subsection{Methodology}
\label{section:datasets}

\heading{Models}
We conduct comprehensive evaluations of \sys{} across two major LMM architectures:
vision-language models (VLMs) and text-to-video (T2V) models (\tableref{table:model-params}).
The VLM architectures integrate ViT-based image encoders (5B and 22B parameters)~\cite{vit-iclr21,vit22b-arxiv23} with language model backbones (Llama3 8B~\cite{llama3-arxiv24}, Qwen2 32B/72B~\cite{qwen2-arxiv24}),
while T2V architectures combine language model encoders with DiT-based diffusion video decoders (5B and 30B parameters)~\cite{moviegen-arxiv24}.
We choose five distinct model combinations ranging from 12B to 94B parameters, as detailed in \tableref{table:model-setup}.

\heading{Datasets}
We employ a combination of diverse open-source datasets~\cite{obelics-arxiv23,laion5b-nips22,scienceqa-nips22,share-gpt4-video-arxiv24,internvid-arxiv23,mmtrail-arxiv24},
comprising pure image-text pairs, interleaved image-text documents, and video-caption pairs.
For VLM models, we adopt the ViT architecture used in Qwen2 VL~\cite{qwen2-vl-arxiv24}. We scale all images to 728px resolution, with each image being encoded by the ViT vision encoder into 169 patch tokens (\texttt{patch\_size}=14, \texttt{spatial\_merge\_size}=4).
Multiple image-text data samples are packed into sequences of 8192 tokens, resulting in a maximum image capacity of $\lfloor 8192/169 \rfloor=48$ per sequence.
For T2V models, we adopt MovieGen's~\cite{moviegen-arxiv24} configuration by transcoding videos to 16 FPS with a maximum duration of 16 seconds.
When processing short videos, we group up to 8 video clips per microbatch while maintaining the total microbatch duration below 16 seconds.

\heading{Baselines}
We benchmark \sys{} against four state-of-the-art baseline systems:

\begin{itemize}
    \item \textbf{Fully Sharded Data Parallel (FSDP)}~\cite{fsdp-arxiv23} is a memory-efficient training strategy standard in PyTorch. It shards model parameters, gradients, and optimizer states across data parallel workers, collecting them via communication only during the computation of the current layer.
    We employ FSDP2 in PyTorch 2.8.0 with ZeRO-3 configuration (\texttt{reshard\_after\_forward=true}) for Transformer blocks to minimize peak memory usage.
    \item \textbf{Megatron-LM}~\cite{megatron-lm-arxiv20} is a widely-used unimodal LLM training framework.
    We use interleaved pipeline parallelism (VPP) and partition LMM layers into model chunks with approximately balanced parameter distribution.
    \item \textbf{nnScaler}~\cite{nnscaler-osdi24}
    automates parallelization for deep neural network training.
    Since generating a single parallelization plan takes several minutes and requires a restart of the training process for plan updates,
    we pre-generate a static parallelization plan before training with a representative training workload.
    For fair cross-framework comparison, we implement nnScaler's model chunk partitioning schemes and memory optimizations in Megatron-LM,
    with performance metrics labeled as ``nnScaler*''.
    \item \textbf{Optimus}~\cite{optimus-atc25}
    proposes coarse-grained and fine-grained bubble scheduling
    to optimize multimodal LLM (MLLM) training with multiple encoders.
    The coarse-grained strategy sequences all modality encoder computations before backbone model execution at the pipeline level,
    while the fine-grained method fills TP communication bubbles with encoder computations
    and is orthogonal to our pipeline design.
    For focused pipeline scheduling comparisons, we implement Optimus' coarse-grained bubble scheduling in \sys{}.
    Due to the lack of support for diffusion decoders, we exclude Optimus from T2V model evaluations.
\end{itemize}

Spindle~\cite{spindle-asplos25} is excluded from the evaluation primarily because it is tailored for static, multi-task scenarios where tasks must be pre-defined prior to training.
This static design fundamentally contrasts with the dynamic and flexible input-handling capabilities required by modern LMMs.

\heading{Testbed}
We conducted large-scale experiments on a cluster of 64 NVIDIA 80GB H800 GPUs distributed across 8 nodes.
Each node is equipped with 128 CPU cores, 256GB of host memory,
and 8 H800 GPUs interconnected via 200 GB/s NVLink.
Inter-node communication relies on an 8$\times$200Gbps RoCEv2 network with a rail-optimized topology.
Additionally, a smaller cluster consisting of two nodes, each populated with 8 NVIDIA 96GB H20 GPUs,
was utilized for comparative baselines against FSDP and Megatron-LM.

\heading{Metrics}
We adopt training iteration time and model FLOPs utilization (MFU) as performance metrics, with all reported values averaged across 10 independent runs.

\subsection{End-to-End Performance}
\label{section:e2e-performance}

\heading{Comparison with LLM Training Systems}
We begin by benchmarking {\sys} against FSDP and Megatron-LM,
both of which are widely adopted frameworks for unimodal LLM training.
These experiments were conducted on the 16-GPU H20 cluster,
leveraging the large 96GB VRAM per GPU to minimize activation recomputation overhead.
As shown in \tableref{table:fsdp-comparison},
FSDP is only 3\% slower than Megatron-LM,
whereas {\sys} achieves a 27\% speedup over Megatron-LM.
Given the marginal performance gap between FSDP and Megatron-LM,
we implement all remaining baselines based on Megatron-LM to ensure a fair cross-framework comparison.

\begin{table}[t]
    \caption{
        VLM-S end-to-end performance of FSDP, Megatron-LM, and {\sys} on 16 H20 GPUs.
    }
    \label{table:fsdp-comparison}

    \small
    \begin{tabular}{>{\raggedright}lccc}
    \toprule
    & \textbf{FSDP}~\cite{fsdp-arxiv23} & \textbf{Megatron-LM}~\cite{megatron-lm-arxiv20} & \textbf{\sys} \\
    \midrule
    Iteration time (s) & 40.270 & 39.053 & 28.606 \\
    Relative time & 1.03 & 1.00 & 0.73 \\
    \bottomrule
    \end{tabular}
\end{table}

\heading{Comparison with Multimodal Training Systems}
We conduct experiments comparing the average end-to-end performance of \sys{} against baseline systems using real datasets and five model configurations summarized in \tableref{table:model-setup}.
As demonstrated in \figref{figure:e2e-perf}a, \sys{} achieves training throughput improvements of 15.6\%--76.2\% over three baseline systems in VLM model setups,
and 36.6\%--97.3\% over two baselines in T2V model configurations,
demonstrating consistent performance gains across diverse model architectures and parameter scales.

\heading{Dynamic Workloads}
To analyze \sys{}'s performance characteristics under dynamic workloads against other baselines,
we investigate the VLM-S model with manual control of image count bounds during training iterations.
We monitor 40 iterations showing two "rise-and-fall" patterns in image counts.
Each pattern consists of:
(1) gradually increasing the lower bound from 0 to 16 while maintaining an upper bound of 32 (iterations 1--5), achieving a peak average of 22 images,
followed by (2) progressively reducing both bounds to zero (iterations 6--20).

\figref{figure:e2e-perf}b reveals \sys{}'s consistent superior performance across all systems.
During high-image-count phases (iterations 1--10),
Megatron-LM suffers significant computational imbalance across modality modules and data microbatches,
exhibiting a 52.9\% slowdown compared to \sys{} at iteration 6.
Although both nnScaler and Optimus partially mitigate dynamic imbalance effects,
they are still 10.4\% slower than \sys{}.
As both image counts and bound gaps decrease (iterations 11--20),
training workloads converge toward pure language tasks,
narrowing the performance gap between \sys{} and baseline systems.
Notably, nnScaler's restriction to 1F1B scheduling mandates all modality modules to be partitioned inside one pipeline segment,
which creates significant pipeline imbalance when image encoder workloads diminish,
resulting in 50.5\% performance degradation during iterations 15--20.

\subsection{Performance Ablation}

\heading{Performance Breakdown}
Using the VLM-S model setup,
we incrementally integrate four key components
(modality-aware partitioner, pipeline stage interleaving, segment reordering, and per-layer memory optimization)
onto Megatron-LM.
As demonstrated in \tableref{table:performance-breakdown},
the modality-aware partitioner alone delivers a 17.3\% performance improvement over the baseline Megatron-LM,
highlighting its critical role in LMM training.
Among optimizations of pipeline schedule searcher,
pipeline stage interleaving provides a substantial 21.6\% performance boost compared to Megatron-LM's default pipelining.
Subsequent segment reordering and per-layer memory optimization enhance performance by an additional 23.9\%
via intelligent reordering of pipeline segments and adaptive memory optimization strategy selection.
In \figref{figure:e2e-perf}b, we visually contrast improvements from modality-aware partitioner and pipeline schedule searcher over dynamic workloads,
separated by the line labeled ``\sys{} (no-opt)'' that excludes pipeline schedule searcher's optimizations.

\begin{table}[t]
    \caption{
        Quantitative impact of \sys{}'s optimizations.
    }
    \label{table:performance-breakdown}

    \small
    \begin{tabular}{>{\raggedright}p{5cm}cc}
    \toprule
    \textbf{Techniques} & \textbf{Iter. Time (s)} & \textbf{$\Delta$\%} \\
    \midrule
    Vanilla Megatron-LM & 26.13 & 0.0\% \\
    \midrule
    + Modality-aware partitioner (\S\ref{section:modality-aware-partitioning}) & 22.27 & 17.3\% \\
    + Pipeline stage interleaving (\S\ref{section:stage-interleaving}) & 18.81 & 38.9\% \\
    + Pipeline segment reordering (\S\ref{section:segment-reordering}) & 17.61 & 48.3\% \\
    + Pre-layer memory optimization (\S\ref{section:memory-optimization}) & 16.05 & 62.8\% \\
    \bottomrule
    \end{tabular}
\end{table}

\heading{Impact of Sub-Microbatch Sizes}
We investigate the influence of modality-specific sub-microbatch sizes on pipeline scheduling and GPU execution efficiency using the VLM-S model,
by testing image sub-microbatch sizes ranging from 4 to 32 and deriving the best and the worst\footnote{We obtain the worst pipeline schedule by inverting the optimization goal of pipeline schedule searcher to maximizing iteration time.} pipeline schedules.

\begin{figure}[!t]
    \includegraphics[width=1\columnwidth]{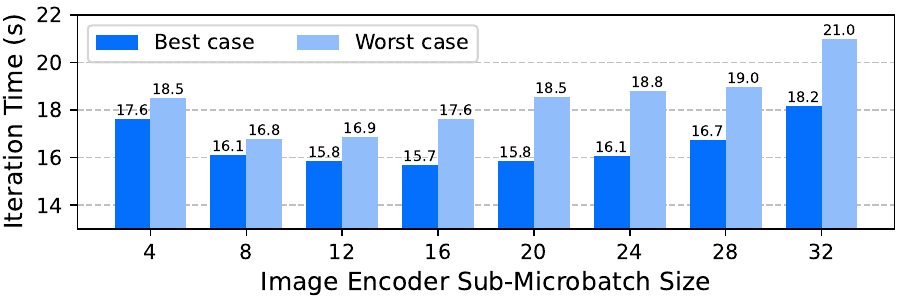}
    \caption{
        Impact of sub-microbatch sizes.
    }
    \label{figure:pipeline-scheduling}
\end{figure}

Our analysis of \figref{figure:pipeline-scheduling} reveals two key findings.
First, smaller sub-microbatch sizes (4--12) significantly reduce the performance gap between best and worst schedules from 15.4\% to 5.1\%.
This narrowed variance indicates reduced sensitivity to schedule configurations,
thereby lowering the difficulty in achieving optimal pipeline schedules.
Second, extremely small sub-microbatch sizes (<8) demonstrate diminishing returns due to underutilized GPU computational capacity,
resulting in a 12.1\% increase in iteration time compared to medium-sized batches.
Through empirical validation, we identify a sub-microbatch size of 12 as the optimal balance point between pipeline schedule flexibility and hardware utilization efficiency.

\begin{figure}[!t]
    \includegraphics[width=1\columnwidth]{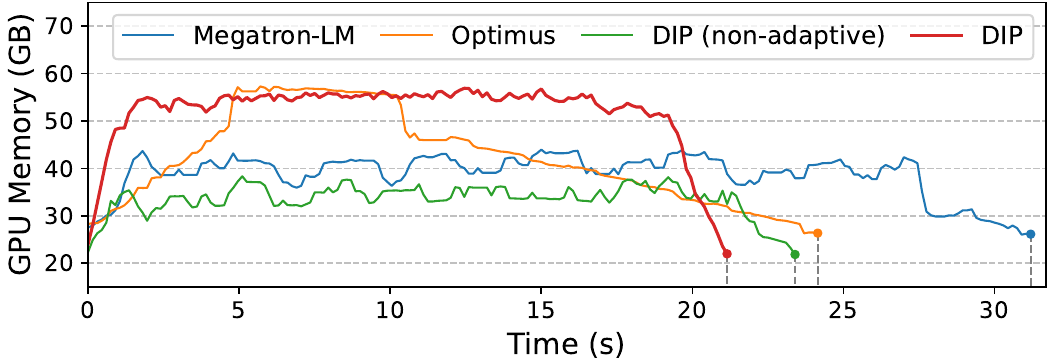}
    \caption{
        Memory usage timelines of the first pipeline rank in VLM-M training.
        The ``\sys{} (non-adaptive)'' disables per-layer memory optimization and does not utilize all available GPU memory.
    }
    \label{figure:adaptive-computation}
\end{figure}

\heading{Impact of Per-Layer Memory Optimization}
We analyze the memory usage timeline of the first pipeline rank during VLM-M model training.
As shown in \figref{figure:adaptive-computation},
baseline systems fail to fully utilize available GPU memory.
Megatron-LM exhibits significant memory fluctuations during the steady 1F1B pipeline phase,
while Optimus manifests a gradual memory increase due to executing all modality encoder computations and storing substantial activation memory prior to the backbone model.
This results in 25.3\% higher peak memory consumption compared to Megatron-LM.
In contrast, \sys{} maintains consistently low memory usage throughout training iterations
by partitioning microbatches into smaller sub-microbatches,
enabling finer-grained interleaving of forward and backward pipeline stages.
Per-layer memory optimization further intelligently selects memory optimization configurations
to achieve full utilization of available GPU memory,
achieving 52.9\% fewer memory fluctuations than Megatron-LM
and delivering 12.2\% performance gains over Optimus under equivalent peak memory conditions.

\begin{figure}[!t]
    \includegraphics[width=1\columnwidth]{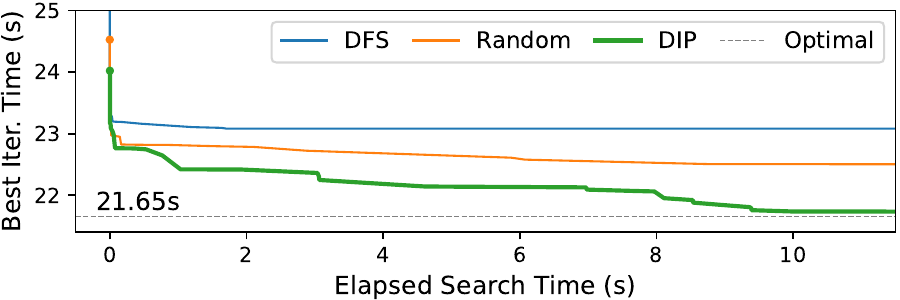}
    \caption{
        Search progress of different exploration strategies.
    }
    \label{figure:search-efficiency}
\end{figure}

\subsection{Planner Evaluation}
\label{section:planner-evaluation}

\heading{Search Efficiency}
To demonstrate the efficiency of \sys{}'s pipeline schedule search,
we track the progression of current best pipeline schedule performance versus elapsed search time,
comparing it with two variants using depth-first search (DFS) and random exploration
instead of MCTS algorithm.
Using the largest VLM-L model as the target workload, we run all search algorithm variants on 64 CPU cores.
\figref{figure:search-efficiency} shows \sys{} achieves near-optimal pipeline schedule performance within 10 seconds,
which can be overlapped with VLM-L's typical 20-second training iteration duration.
In contrast, DFS and random exploration strategies fail to quickly identify optimal pipeline schedules
due to their lack of guided search optimization with performance scores like MCTS.

\begin{figure}[!t]
    \includegraphics[width=1\columnwidth]{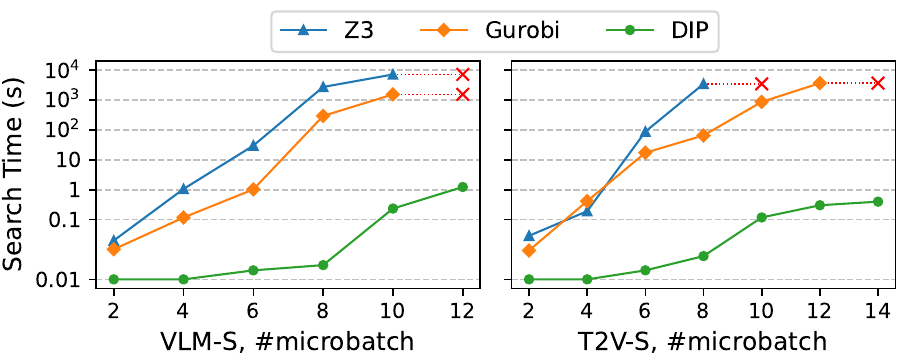}
    \caption{
        Search time comparison of Z3, Gurobi, and {\sys} across varying microbatch counts for VLM-S and T2V-S.
        The Y-axis is plotted on a log scale.
        Red crosses (\textcolor{red}{$\times$}) indicate timeouts (>3 hours).
    }
    \label{figure:search-scalability}
\end{figure}

\heading{Search Scalability}
We evaluate search scalability using two model configurations: VLM-S and T2V-S.
Specifically, we measure the search latency of {\sys} as the number of microbatches increases,
which correspondingly increases the number of pipeline stages and the complexity of the pipeline schedule.
We benchmark {\sys} against two state-of-the-art SMT/ILP solvers: Z3~\cite{z3} and Gurobi~\cite{gurobi} (version 13.0).
For the experimental setup, both {\sys} and Gurobi utilize up to 64 CPU threads,
whereas Z3 operates on a single thread as it inherently lacks support for multi-threaded solving.
As shown in \figref{figure:search-scalability}, {\sys} consistently maintains a search time under 10 seconds.
In contrast, the search times for both Z3 and Gurobi exhibit exponential growth with respect to the number of microbatches.
Consequently, both solvers cannot complete in 30 minutes when the number of microbatches surpasses 10,
rendering them impractical for dynamic pipeline schedule generation in LMM training.

\begin{figure}[!t]
    \includegraphics[width=1\columnwidth]{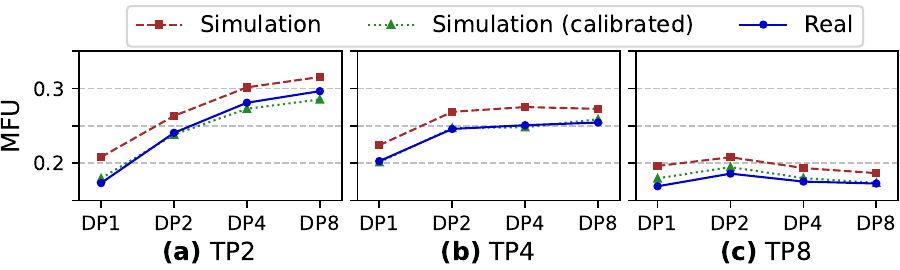}
    \caption{
        Comparison between pre- and post-calibration simulation results versus actual GPU executions.
    }
    \label{figure:simulation-accuracy}
\end{figure}

\heading{Simulation Accuracy}
We assess the accuracy of \sys{}'s training simulator through a grid-search experiment for VLM-M across 64 GPUs,
and compare simulated results against actual GPU executions.
We systematically evaluate all valid combinations of DP, PP, and TP sizes where all values are powers of two and TP $\leq 8$.
As illustrated in \figref{figure:simulation-accuracy},
the optimal parallelism configuration is DP8, TP2, and PP4, achieving a MFU of 29.7\%.
Although \sys{}'s default simulation settings initially exhibit relative errors up to 10\% compared to ground-truth measurements,
the simulator still successfully predicts the optimal parallel configuration.
Through calibration via offline microbenchmarks that align efficiency scaling factors for matrix multiplications and collective communication operations,
the training simulator achieves an average simulation accuracy of 97.6\%.

\subsection{Large-Scale Simulation}

To validate \sys{}'s effectiveness in large-scale H100 GPU cluster environments,
we conducted two experiments using the training simulator with substantial large multimodal models (VLM-XL and T2V-XL)
to evaluate its theoretical improvements over baseline approaches.
Detailed configurations for both models and GPU clusters are presented in \tableref{table:xl-model-setup}.

\begin{figure}[t]
    \includegraphics[width=1\columnwidth]{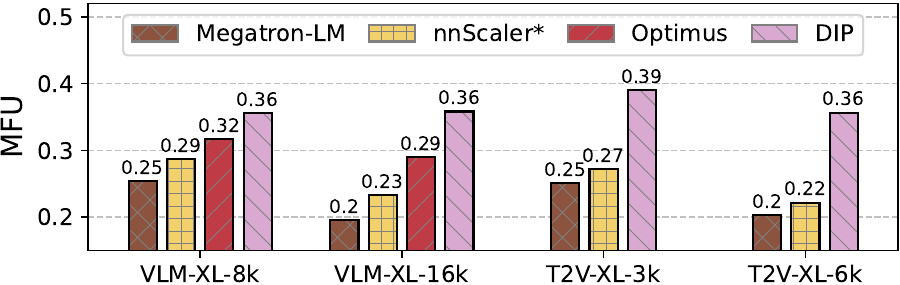}
    \caption{
        Large scale simulations on H100 clusters.
    }
    \label{figure:large-scale-simulation}
\end{figure}

\heading{Results}
Experimental results in \figref{figure:large-scale-simulation} demonstrate that \sys{} achieves MFU scores of 0.36 for VLM-XL and 0.39 for T2V-XL.
The system consistently outperforms baseline methods by up to 82.8\%,
particularly with larger pipeline parallelism sizes,
as larger PP dimensions introduce more complex pipeline structures
that require meticulous orchestration between stages by the pipeline schedule searcher.

\section{Related Work}

\heading{Automated Training Parallelization}
Several systems have been proposed to automate parallelization strategies for training deep learning models~\cite{alpa-osdi22,nnscaler-osdi24,unity-osdi22}. These systems perform full-scale model planning across DP, PP, and TP dimensions. Although they employ exhaustive search algorithms to generate near-optimal training configurations, the planning process often incurs substantial time overhead—requiring several minutes to complete~\cite{nnscaler-osdi24}. This makes such approaches impractical for LMM training, which demands dynamic and adaptive pipeline scheduling.

\heading{Training Systems for Multimodal Models}
Numerous system optimizations~\cite{distmm-nsdi24,optimus-atc25,disttrain-arxiv24,spindle-asplos25,orch-mllm-arxiv25,cornstarch-arxiv25,diffusionpipe-mlsys24} have been developed to address architectural and training data heterogeneity in multimodal model training. For instance, DistMM~\cite{distmm-nsdi24} tackles model and data heterogeneity in traditional CLIP-based models, where all encoders are parallel. However, this design does not generalize to LMMs, which exhibit sequential dependencies between encoders, backbones and decoders. Similarly, both Spindle~\cite{spindle-asplos25} and Optimus~\cite{optimus-atc25} target multimodal LLMs under predefined training tasks, overlooking the inherent dynamicity in LMM training pipelines.
Specifically, Spindle is designed for a static, multi-task setting where all tasks are pre-defined before training starts. This contrasts with the dynamic and flexible input-handling capabilities required by modern LMMs, which is the focus of \sys{}.

\heading{Pipeline Parallelism Scheduling}
Beyond Megatron-LM's 1F1B schedule, several alternative pipeline schedules have been proposed~\cite{chimera-sc21,hanayo-sc23,zero-bubble-iclr24,graph-pipe-asplos25}. Chimera~\cite{chimera-sc21} introduces bidirectional pipelines to reduce pipeline bubbles, while zero bubble pipeline~\cite{zero-bubble-iclr24} further minimizes bubbles by decoupling backward computations into input gradient and weight gradient phases. However, these methods assume fixed pipeline stage latencies and still suffer from the dynamic imbalance problem. GraphPipe~\cite{graph-pipe-asplos25} targets models with heterogeneous modules by organizing computations into a directed acyclic graph (DAG) rather than sequential stages, enabling more flexible scheduling. \sys{}'s pipeline schedule searcher can be extended to incorporate such custom schedules like GraphPipe.

\heading{Pipeline with Variable Sequence Lengths}
Several systems address text sequence length variations in unimodal LLM training~\cite{wlb-llm-osdi25,dynapipe-eurosys24,flexpipe-atc25,byte-transformer-ipdps23,sppo-arxiv25}. For example, DynaPipe~\cite{dynapipe-eurosys24} optimizes micro-batch construction using dynamic programming for multi-task training. WLB-LLM~\cite{wlb-llm-osdi25} proposes a fine-grained per-document sharding strategy to balance workloads in context parallelism groups. FlexPipe~\cite{flexpipe-atc25} dynamically adjusts pipeline size via a live flexibility mechanism to reassign GPU workers processing short sequences to assist with longer ones. While these approaches are effective for text-based models, they do not handle multi-modal data. In contrast, our approach supports multiple modalities within a single pipeline.

\begin{table}[!t]
    \caption{
        Model combinations used in the large-scale simulation.
    }
    \label{table:xl-model-setup}

    \footnotesize
    \begin{tabular}{llcccc}
    \toprule
    \textbf{Name} & \textbf{Model Setup} & \textbf{DP} & \textbf{TP} & \textbf{PP} & \textbf{\#GPU} \\
    \midrule
    VLM-XL-8k & ViT 22B + GPT 175B & 128 & 8 & 8 & 8192 \\
    VLM-XL-16k & ViT 22B + GPT 175B & 128 & 8 & 16 & 16384 \\
    \midrule
    T2V-XL-3k & Qwen2 72B + DiT 30B & 96 & 8 & 4 & 3072 \\
    T2V-XL-6k & Qwen2 72B + DiT 30B & 96 & 8 & 8 & 6144 \\
    \bottomrule
    \end{tabular}
\end{table}

\section{Conclusion}

Efficient large multimodal model training is challenging due to pipeline stage imbalance and training data dynamicity.
In this paper, we propose {\sys} to address dynamic imbalance of LMM training
with adaptive modality-aware partitioning and efficient pipeline schedule search.
Our experimental results demonstrate that {\sys} outperforms existing state-of-the-art training systems by up to 97.3\% in training throughput.

\begin{acks}
We sincerely thank our shepherd Jongsoo Park, and the anonymous reviewers for their insightful suggestions.
This work was partially supported by NSFC (No. 62372287 and U24A20235).
Zeyu Mi (\href{mailto:yzmizeyu@sjtu.edu.cn}{yzmizeyu@sjtu.edu.cn}) is the corresponding author.
\end{acks}


\bibliographystyle{ACM-Reference-Format}
\balance
\bibliography{paper}

@inproceedings{alpa-osdi22,
  author = {Lianmin Zheng and Zhuohan Li and Hao Zhang and Yonghao Zhuang and Zhifeng Chen and Yanping Huang and Yida Wang and Yuanzhong Xu and Danyang Zhuo and Eric P. Xing and Joseph E. Gonzalez and Ion Stoica},
  title = {{Alpa}: Automating Inter- and Intra-Operator Parallelism for Distributed Deep Learning},
  booktitle = {16th USENIX Symposium on Operating Systems Design and Implementation (OSDI 22)},
  year = {2022},
}

@inproceedings{nnscaler-osdi24,
  author = {Zhiqi Lin and Youshan Miao and Quanlu Zhang and Fan Yang and Yi Zhu and Cheng Li and Saeed Maleki and Xu Cao and Ning Shang and Yilei Yang and Weijiang Xu and Mao Yang and Lintao Zhang and Lidong Zhou},
  title = {{nnScaler}: Constraint-Guided Parallelization Plan Generation for Deep Learning Training},
  booktitle = {18th USENIX Symposium on Operating Systems Design and Implementation (OSDI 24)},
  year = {2024},
}

@inproceedings{wlb-llm-osdi25,
  author = {Zheng Wang and Anna Cai and Xinfeng Xie and Zaifeng Pan and Yue Guan and Weiwei Chu and Jie Wang and Shikai Li and Jianyu Huang and Chris Cai and Yuchen Hao and Yufei Ding},
  title = {{WLB-LLM}: Workload-Balanced {4D} Parallelism for Large Language Model Training},
  booktitle = {19th USENIX Symposium on Operating Systems Design and Implementation (OSDI 25)},
  year = {2025},
}

@inproceedings {unity-osdi22,
  author = {Colin Unger and Zhihao Jia and Wei Wu and Sina Lin and Mandeep Baines and Carlos Efrain Quintero Narvaez and Vinay Ramakrishnaiah and Nirmal Prajapati and Pat McCormick and Jamaludin Mohd-Yusof and Xi Luo and Dheevatsa Mudigere and Jongsoo Park and Misha Smelyanskiy and Alex Aiken},
  title = {{Unity}: Accelerating {DNN} Training Through Joint Optimization of Algebraic Transformations and Parallelization},
  booktitle = {16th USENIX Symposium on Operating Systems Design and Implementation (OSDI 22)},
  year = {2022},
}

@inproceedings{optimus-atc25,
  title={{Optimus}: Accelerating Large-Scale Multi-Modal {LLM} Training by Bubble Exploitation},
  author={Weiqi Feng and Yangrui Chen and Shaoyu Wang and Yanghua Peng and Haibin Lin and Minlan Yu},
booktitle = {2025 USENIX Annual Technical Conference (USENIX ATC 25)},
year = {2025},
address = {Boston, MA, USA},
pages = {161--178},
publisher = {USENIX Association},
month = jul
}

@inproceedings{flexpipe-atc25,
  title={{FlexPipe}: Maximizing Training Efficiency for Transformer-Based Models with Variable-Length Inputs},
  author={Hairui Zhao and Qi Tian and Hongliang Li and Zizhong Chen},
booktitle = {2025 USENIX Annual Technical Conference (USENIX ATC 25)},
year = {2025},
address = {Boston, MA, USA},
pages = {143--159},
publisher = {USENIX Association},
month = jul
}

@misc{disttrain-arxiv24,
  howpublished={arXiv},
  title={{DistTrain}: Addressing Model and Data Heterogeneity with Disaggregated Training for Multimodal Large Language Models},
  author={Zili Zhang and Yinmin Zhong and Ranchen Ming and Hanpeng Hu and Jianjian Sun and Zheng Ge and Yibo Zhu and Xin Jin},
  year={2024},
  eprint={2408.04275},
  archivePrefix={arXiv},
  primaryClass={cs.DC},
  url={https://arxiv.org/abs/2408.04275},
}

@inproceedings{diffusionpipe-mlsys24,
  author = {Ye Tian and Zhen Jia and Ziyue Luo and Yida Wang and Chuan Wu},
  booktitle = {Proceedings of Machine Learning and Systems},
  title = {{DiffusionPipe}: Training Large Diffusion Models with Efficient Pipelines},
  year = {2024}
}

@misc{megatron-lm-arxiv20,
  howpublished={arXiv},
  title={{Megatron-LM}: Training Multi-Billion Parameter Language Models Using Model Parallelism},
  author={Mohammad Shoeybi and Mostofa Patwary and Raul Puri and Patrick LeGresley and Jared Casper and Bryan Catanzaro},
  year={2020},
  eprint={1909.08053},
  archivePrefix={arXiv},
  primaryClass={cs.CL},
  url={https://arxiv.org/abs/1909.08053},
}

@inproceedings{vit-iclr21,
title={An Image is Worth 16x16 Words: Transformers for Image Recognition at Scale},
author={Alexey Dosovitskiy and Lucas Beyer and Alexander Kolesnikov and Dirk Weissenborn and Xiaohua Zhai and Thomas Unterthiner and Mostafa Dehghani and Matthias Minderer and Georg Heigold and Sylvain Gelly and Jakob Uszkoreit and Neil Houlsby},
booktitle={International Conference on Learning Representations},
year={2021},
url={https://openreview.net/forum?id=YicbFdNTTy}
}

@misc{llama3-arxiv24,
      title={The Llama 3 Herd of Models},
      author={Aaron Grattafiori and Abhimanyu Dubey and Abhinav Jauhri and Abhinav Pandey and Abhishek Kadian and Ahmad Al-Dahle and Aiesha Letman and Akhil Mathur and Alan Schelten and Alex Vaughan and Amy Yang and Angela Fan and Anirudh Goyal and Anthony Hartshorn and Aobo Yang and Archi Mitra and Archie Sravankumar and Artem Korenev and Arthur Hinsvark and Arun Rao and Aston Zhang and Aurelien Rodriguez and Austen Gregerson and Ava Spataru and Baptiste Roziere and Bethany Biron and Binh Tang and Bobbie Chern and Charlotte Caucheteux and Chaya Nayak and Chloe Bi and Chris Marra and Chris McConnell and Christian Keller and Christophe Touret and Chunyang Wu and Corinne Wong and Cristian Canton Ferrer and Cyrus Nikolaidis and Damien Allonsius and Daniel Song and Danielle Pintz and Danny Livshits and Danny Wyatt and David Esiobu and Dhruv Choudhary and Dhruv Mahajan and Diego Garcia-Olano and Diego Perino and Dieuwke Hupkes and Egor Lakomkin and Ehab AlBadawy and Elina Lobanova and Emily Dinan and Eric Michael Smith and Filip Radenovic and Francisco Guzmán and Frank Zhang and Gabriel Synnaeve and Gabrielle Lee and Georgia Lewis Anderson and Govind Thattai and Graeme Nail and Gregoire Mialon and Guan Pang and Guillem Cucurell and Hailey Nguyen and Hannah Korevaar and Hu Xu and Hugo Touvron and Iliyan Zarov and Imanol Arrieta Ibarra and Isabel Kloumann and Ishan Misra and Ivan Evtimov and Jack Zhang and Jade Copet and Jaewon Lee and Jan Geffert and Jana Vranes and Jason Park and Jay Mahadeokar and Jeet Shah and Jelmer van der Linde and Jennifer Billock and Jenny Hong and Jenya Lee and Jeremy Fu and Jianfeng Chi and Jianyu Huang and Jiawen Liu and Jie Wang and Jiecao Yu and Joanna Bitton and Joe Spisak and Jongsoo Park and Joseph Rocca and Joshua Johnstun and Joshua Saxe and Junteng Jia and Kalyan Vasuden Alwala and Karthik Prasad and Kartikeya Upasani and Kate Plawiak and Ke Li and Kenneth Heafield and Kevin Stone and Khalid El-Arini and Krithika Iyer and Kshitiz Malik and Kuenley Chiu and Kunal Bhalla and Kushal Lakhotia and Lauren Rantala-Yeary and Laurens van der Maaten and Lawrence Chen and Liang Tan and Liz Jenkins and Louis Martin and Lovish Madaan and Lubo Malo and Lukas Blecher and Lukas Landzaat and Luke de Oliveira and Madeline Muzzi and Mahesh Pasupuleti and Mannat Singh and Manohar Paluri and Marcin Kardas and Maria Tsimpoukelli and Mathew Oldham and Mathieu Rita and Maya Pavlova and Melanie Kambadur and Mike Lewis and Min Si and Mitesh Kumar Singh and Mona Hassan and Naman Goyal and Narjes Torabi and Nikolay Bashlykov and Nikolay Bogoychev and Niladri Chatterji and Ning Zhang and Olivier Duchenne and Onur Çelebi and Patrick Alrassy and Pengchuan Zhang and Pengwei Li and Petar Vasic and Peter Weng and Prajjwal Bhargava and Pratik Dubal and Praveen Krishnan and Punit Singh Koura and Puxin Xu and Qing He and Qingxiao Dong and Ragavan Srinivasan and Raj Ganapathy and Ramon Calderer and Ricardo Silveira Cabral and Robert Stojnic and Roberta Raileanu and Rohan Maheswari and Rohit Girdhar and Rohit Patel and Romain Sauvestre and Ronnie Polidoro and Roshan Sumbaly and Ross Taylor and Ruan Silva and Rui Hou and Rui Wang and Saghar Hosseini and Sahana Chennabasappa and Sanjay Singh and Sean Bell and Seohyun Sonia Kim and Sergey Edunov and Shaoliang Nie and Sharan Narang and Sharath Raparthy and Sheng Shen and Shengye Wan and Shruti Bhosale and Shun Zhang and Simon Vandenhende and Soumya Batra and Spencer Whitman and Sten Sootla and Stephane Collot and Suchin Gururangan and Sydney Borodinsky and Tamar Herman and Tara Fowler and Tarek Sheasha and Thomas Georgiou and Thomas Scialom and Tobias Speckbacher and Todor Mihaylov and Tong Xiao and Ujjwal Karn and Vedanuj Goswami and Vibhor Gupta and Vignesh Ramanathan and Viktor Kerkez and Vincent Gonguet and Virginie Do and Vish Vogeti and Vítor Albiero and Vladan Petrovic and Weiwei Chu and Wenhan Xiong and Wenyin Fu and Whitney Meers and Xavier Martinet and Xiaodong Wang and Xiaofang Wang and Xiaoqing Ellen Tan and Xide Xia and Xinfeng Xie and Xuchao Jia and Xuewei Wang and Yaelle Goldschlag and Yashesh Gaur and Yasmine Babaei and Yi Wen and Yiwen Song and Yuchen Zhang and Yue Li and Yuning Mao and Zacharie Delpierre Coudert and Zheng Yan and Zhengxing Chen and Zoe Papakipos and Aaditya Singh and Aayushi Srivastava and Abha Jain and Adam Kelsey and Adam Shajnfeld and Adithya Gangidi and Adolfo Victoria and Ahuva Goldstand and Ajay Menon and Ajay Sharma and Alex Boesenberg and Alexei Baevski and Allie Feinstein and Amanda Kallet and Amit Sangani and Amos Teo and Anam Yunus and Andrei Lupu and Andres Alvarado and Andrew Caples and Andrew Gu and Andrew Ho and Andrew Poulton and Andrew Ryan and Ankit Ramchandani and Annie Dong and Annie Franco and Anuj Goyal and Aparajita Saraf and Arkabandhu Chowdhury and Ashley Gabriel and Ashwin Bharambe and Assaf Eisenman and Azadeh Yazdan and Beau James and Ben Maurer and Benjamin Leonhardi and Bernie Huang and Beth Loyd and Beto De Paola and Bhargavi Paranjape and Bing Liu and Bo Wu and Boyu Ni and Braden Hancock and Bram Wasti and Brandon Spence and Brani Stojkovic and Brian Gamido and Britt Montalvo and Carl Parker and Carly Burton and Catalina Mejia and Ce Liu and Changhan Wang and Changkyu Kim and Chao Zhou and Chester Hu and Ching-Hsiang Chu and Chris Cai and Chris Tindal and Christoph Feichtenhofer and Cynthia Gao and Damon Civin and Dana Beaty and Daniel Kreymer and Daniel Li and David Adkins and David Xu and Davide Testuggine and Delia David and Devi Parikh and Diana Liskovich and Didem Foss and Dingkang Wang and Duc Le and Dustin Holland and Edward Dowling and Eissa Jamil and Elaine Montgomery and Eleonora Presani and Emily Hahn and Emily Wood and Eric-Tuan Le and Erik Brinkman and Esteban Arcaute and Evan Dunbar and Evan Smothers and Fei Sun and Felix Kreuk and Feng Tian and Filippos Kokkinos and Firat Ozgenel and Francesco Caggioni and Frank Kanayet and Frank Seide and Gabriela Medina Florez and Gabriella Schwarz and Gada Badeer and Georgia Swee and Gil Halpern and Grant Herman and Grigory Sizov and Guangyi and Zhang and Guna Lakshminarayanan and Hakan Inan and Hamid Shojanazeri and Han Zou and Hannah Wang and Hanwen Zha and Haroun Habeeb and Harrison Rudolph and Helen Suk and Henry Aspegren and Hunter Goldman and Hongyuan Zhan and Ibrahim Damlaj and Igor Molybog and Igor Tufanov and Ilias Leontiadis and Irina-Elena Veliche and Itai Gat and Jake Weissman and James Geboski and James Kohli and Janice Lam and Japhet Asher and Jean-Baptiste Gaya and Jeff Marcus and Jeff Tang and Jennifer Chan and Jenny Zhen and Jeremy Reizenstein and Jeremy Teboul and Jessica Zhong and Jian Jin and Jingyi Yang and Joe Cummings and Jon Carvill and Jon Shepard and Jonathan McPhie and Jonathan Torres and Josh Ginsburg and Junjie Wang and Kai Wu and Kam Hou U and Karan Saxena and Kartikay Khandelwal and Katayoun Zand and Kathy Matosich and Kaushik Veeraraghavan and Kelly Michelena and Keqian Li and Kiran Jagadeesh and Kun Huang and Kunal Chawla and Kyle Huang and Lailin Chen and Lakshya Garg and Lavender A and Leandro Silva and Lee Bell and Lei Zhang and Liangpeng Guo and Licheng Yu and Liron Moshkovich and Luca Wehrstedt and Madian Khabsa and Manav Avalani and Manish Bhatt and Martynas Mankus and Matan Hasson and Matthew Lennie and Matthias Reso and Maxim Groshev and Maxim Naumov and Maya Lathi and Meghan Keneally and Miao Liu and Michael L. Seltzer and Michal Valko and Michelle Restrepo and Mihir Patel and Mik Vyatskov and Mikayel Samvelyan and Mike Clark and Mike Macey and Mike Wang and Miquel Jubert Hermoso and Mo Metanat and Mohammad Rastegari and Munish Bansal and Nandhini Santhanam and Natascha Parks and Natasha White and Navyata Bawa and Nayan Singhal and Nick Egebo and Nicolas Usunier and Nikhil Mehta and Nikolay Pavlovich Laptev and Ning Dong and Norman Cheng and Oleg Chernoguz and Olivia Hart and Omkar Salpekar and Ozlem Kalinli and Parkin Kent and Parth Parekh and Paul Saab and Pavan Balaji and Pedro Rittner and Philip Bontrager and Pierre Roux and Piotr Dollar and Polina Zvyagina and Prashant Ratanchandani and Pritish Yuvraj and Qian Liang and Rachad Alao and Rachel Rodriguez and Rafi Ayub and Raghotham Murthy and Raghu Nayani and Rahul Mitra and Rangaprabhu Parthasarathy and Raymond Li and Rebekkah Hogan and Robin Battey and Rocky Wang and Russ Howes and Ruty Rinott and Sachin Mehta and Sachin Siby and Sai Jayesh Bondu and Samyak Datta and Sara Chugh and Sara Hunt and Sargun Dhillon and Sasha Sidorov and Satadru Pan and Saurabh Mahajan and Saurabh Verma and Seiji Yamamoto and Sharadh Ramaswamy and Shaun Lindsay and Shaun Lindsay and Sheng Feng and Shenghao Lin and Shengxin Cindy Zha and Shishir Patil and Shiva Shankar and Shuqiang Zhang and Shuqiang Zhang and Sinong Wang and Sneha Agarwal and Soji Sajuyigbe and Soumith Chintala and Stephanie Max and Stephen Chen and Steve Kehoe and Steve Satterfield and Sudarshan Govindaprasad and Sumit Gupta and Summer Deng and Sungmin Cho and Sunny Virk and Suraj Subramanian and Sy Choudhury and Sydney Goldman and Tal Remez and Tamar Glaser and Tamara Best and Thilo Koehler and Thomas Robinson and Tianhe Li and Tianjun Zhang and Tim Matthews and Timothy Chou and Tzook Shaked and Varun Vontimitta and Victoria Ajayi and Victoria Montanez and Vijai Mohan and Vinay Satish Kumar and Vishal Mangla and Vlad Ionescu and Vlad Poenaru and Vlad Tiberiu Mihailescu and Vladimir Ivanov and Wei Li and Wenchen Wang and Wenwen Jiang and Wes Bouaziz and Will Constable and Xiaocheng Tang and Xiaojian Wu and Xiaolan Wang and Xilun Wu and Xinbo Gao and Yaniv Kleinman and Yanjun Chen and Ye Hu and Ye Jia and Ye Qi and Yenda Li and Yilin Zhang and Ying Zhang and Yossi Adi and Youngjin Nam and Yu and Wang and Yu Zhao and Yuchen Hao and Yundi Qian and Yunlu Li and Yuzi He and Zach Rait and Zachary DeVito and Zef Rosnbrick and Zhaoduo Wen and Zhenyu Yang and Zhiwei Zhao and Zhiyu Ma},
      year={2024},
      eprint={2407.21783},
      archivePrefix={arXiv},
      primaryClass={cs.AI},
      url={https://arxiv.org/abs/2407.21783},
}

@misc{qwen2-arxiv24,
  howpublished={arXiv},
  title={{Qwen2} Technical Report},
  author={An Yang and Baosong Yang and Binyuan Hui and Bo Zheng and Bowen Yu and Chang Zhou and Chengpeng Li and Chengyuan Li and Dayiheng Liu and Fei Huang and Guanting Dong and Haoran Wei and Huan Lin and Jialong Tang and Jialin Wang and Jian Yang and Jianhong Tu and Jianwei Zhang and Jianxin Ma and Jianxin Yang and Jin Xu and Jingren Zhou and Jinze Bai and Jinzheng He and Junyang Lin and Kai Dang and Keming Lu and Keqin Chen and Kexin Yang and Mei Li and Mingfeng Xue and Na Ni and Pei Zhang and Peng Wang and Ru Peng and Rui Men and Ruize Gao and Runji Lin and Shijie Wang and Shuai Bai and Sinan Tan and Tianhang Zhu and Tianhao Li and Tianyu Liu and Wenbin Ge and Xiaodong Deng and Xiaohuan Zhou and Xingzhang Ren and Xinyu Zhang and Xipin Wei and Xuancheng Ren and Xuejing Liu and Yang Fan and Yang Yao and Yichang Zhang and Yu Wan and Yunfei Chu and Yuqiong Liu and Zeyu Cui and Zhenru Zhang and Zhifang Guo and Zhihao Fan},
  year={2024},
  eprint={2407.10671},
  archivePrefix={arXiv},
  primaryClass={cs.CL},
  url={https://arxiv.org/abs/2407.10671},
}

@misc{qwen2-5-arxiv25,
  howpublished={arXiv},
      title={{Qwen2.5} Technical Report},
      author={An Yang and Baosong Yang and Beichen Zhang and Binyuan Hui and Bo Zheng and Bowen Yu and Chengyuan Li and Dayiheng Liu and Fei Huang and Haoran Wei and Huan Lin and Jian Yang and Jianhong Tu and Jianwei Zhang and Jianxin Yang and Jiaxi Yang and Jingren Zhou and Junyang Lin and Kai Dang and Keming Lu and Keqin Bao and Kexin Yang and Le Yu and Mei Li and Mingfeng Xue and Pei Zhang and Qin Zhu and Rui Men and Runji Lin and Tianhao Li and Tianyi Tang and Tingyu Xia and Xingzhang Ren and Xuancheng Ren and Yang Fan and Yang Su and Yichang Zhang and Yu Wan and Yuqiong Liu and Zeyu Cui and Zhenru Zhang and Zihan Qiu},
      year={2025},
      eprint={2412.15115},
      archivePrefix={arXiv},
      primaryClass={cs.CL},
      url={https://arxiv.org/abs/2412.15115},
}

@misc{vit22b-arxiv23,
  howpublished={arXiv},
  title={Scaling Vision Transformers to 22 Billion Parameters},
  author={Mostafa Dehghani and Josip Djolonga and Basil Mustafa and Piotr Padlewski and Jonathan Heek and Justin Gilmer and Andreas Steiner and Mathilde Caron and Robert Geirhos and Ibrahim Alabdulmohsin and Rodolphe Jenatton and Lucas Beyer and Michael Tschannen and Anurag Arnab and Xiao Wang and Carlos Riquelme and Matthias Minderer and Joan Puigcerver and Utku Evci and Manoj Kumar and Sjoerd van Steenkiste and Gamaleldin F. Elsayed and Aravindh Mahendran and Fisher Yu and Avital Oliver and Fantine Huot and Jasmijn Bastings and Mark Patrick Collier and Alexey Gritsenko and Vighnesh Birodkar and Cristina Vasconcelos and Yi Tay and Thomas Mensink and Alexander Kolesnikov and Filip Pavetić and Dustin Tran and Thomas Kipf and Mario Lučić and Xiaohua Zhai and Daniel Keysers and Jeremiah Harmsen and Neil Houlsby},
  year={2023},
  eprint={2302.05442},
  archivePrefix={arXiv},
  primaryClass={cs.CV},
  url={https://arxiv.org/abs/2302.05442},
}

@misc{z3,
  title        = {The {Z3} Theorem Prover},
  author       = {Z3 developers},
  year={2025},
  howpublished = {\url{https://github.com/Z3Prover/z3}},
}

@InProceedings{mcts,
author="Coulom, R{\'e}mi",
editor="van den Herik, H. Jaap
and Ciancarini, Paolo
and Donkers, H. H. L. M. (Jeroen)",
title="Efficient Selectivity and Backup Operators in {Monte-Carlo} Tree Search",
booktitle="Computers and Games",
year="2007",
publisher="Springer Berlin Heidelberg",
address="Berlin, Heidelberg",
pages="72--83",
isbn="978-3-540-75538-8"
}

@misc{moviegen-arxiv24,
  howpublished={arXiv},
  title={{Movie Gen}: A Cast of Media Foundation Models},
  author={Adam Polyak and Amit Zohar and Andrew Brown and Andros Tjandra and Animesh Sinha and Ann Lee and Apoorv Vyas and Bowen Shi and Chih-Yao Ma and Ching-Yao Chuang and David Yan and Dhruv Choudhary and Dingkang Wang and Geet Sethi and Guan Pang and Haoyu Ma and Ishan Misra and Ji Hou and Jialiang Wang and Kiran Jagadeesh and Kunpeng Li and Luxin Zhang and Mannat Singh and Mary Williamson and Matt Le and Matthew Yu and Mitesh Kumar Singh and Peizhao Zhang and Peter Vajda and Quentin Duval and Rohit Girdhar and Roshan Sumbaly and Sai Saketh Rambhatla and Sam Tsai and Samaneh Azadi and Samyak Datta and Sanyuan Chen and Sean Bell and Sharadh Ramaswamy and Shelly Sheynin and Siddharth Bhattacharya and Simran Motwani and Tao Xu and Tianhe Li and Tingbo Hou and Wei-Ning Hsu and Xi Yin and Xiaoliang Dai and Yaniv Taigman and Yaqiao Luo and Yen-Cheng Liu and Yi-Chiao Wu and Yue Zhao and Yuval Kirstain and Zecheng He and Zijian He and Albert Pumarola and Ali Thabet and Artsiom Sanakoyeu and Arun Mallya and Baishan Guo and Boris Araya and Breena Kerr and Carleigh Wood and Ce Liu and Cen Peng and Dimitry Vengertsev and Edgar Schonfeld and Elliot Blanchard and Felix Juefei-Xu and Fraylie Nord and Jeff Liang and John Hoffman and Jonas Kohler and Kaolin Fire and Karthik Sivakumar and Lawrence Chen and Licheng Yu and Luya Gao and Markos Georgopoulos and Rashel Moritz and Sara K. Sampson and Shikai Li and Simone Parmeggiani and Steve Fine and Tara Fowler and Vladan Petrovic and Yuming Du},
  year={2024},
  eprint={2410.13720},
  archivePrefix={arXiv},
  primaryClass={cs.CV},
  url={https://arxiv.org/abs/2410.13720},
}

@inproceedings{laion5b-nips22,
  author = {Schuhmann, Christoph and Beaumont, Romain and Vencu, Richard and Gordon, Cade and Wightman, Ross and Cherti, Mehdi and Coombes, Theo and Katta, Aarush and Mullis, Clayton and Wortsman, Mitchell and Schramowski, Patrick and Kundurthy, Srivatsa and Crowson, Katherine and Schmidt, Ludwig and Kaczmarczyk, Robert and Jitsev, Jenia},
  title = {{LAION-5B}: An Open Large-Scale Dataset for Training Next Generation Image-Text Models},
  year = {2022},
  isbn = {9781713871088},
  publisher = {Curran Associates Inc.},
  address = {Red Hook, NY, USA},
  booktitle = {Proceedings of the 36th International Conference on Neural Information Processing Systems},
  articleno = {1833},
  numpages = {17},
  location = {New Orleans, LA, USA},
  series = {NIPS '22}
}

@misc{step-t2v-arxiv25,
  howpublished={arXiv},
  title={{Step-Video-T2V} Technical Report: The Practice, Challenges, and Future of Video Foundation Model},
  author={Guoqing Ma and Haoyang Huang and Kun Yan and Liangyu Chen and Nan Duan and Shengming Yin and Changyi Wan and Ranchen Ming and Xiaoniu Song and Xing Chen and Yu Zhou and Deshan Sun and Deyu Zhou and Jian Zhou and Kaijun Tan and Kang An and Mei Chen and Wei Ji and Qiling Wu and Wen Sun and Xin Han and Yanan Wei and Zheng Ge and Aojie Li and Bin Wang and Bizhu Huang and Bo Wang and Brian Li and Changxing Miao and Chen Xu and Chenfei Wu and Chenguang Yu and Dapeng Shi and Dingyuan Hu and Enle Liu and Gang Yu and Ge Yang and Guanzhe Huang and Gulin Yan and Haiyang Feng and Hao Nie and Haonan Jia and Hanpeng Hu and Hanqi Chen and Haolong Yan and Heng Wang and Hongcheng Guo and Huilin Xiong and Huixin Xiong and Jiahao Gong and Jianchang Wu and Jiaoren Wu and Jie Wu and Jie Yang and Jiashuai Liu and Jiashuo Li and Jingyang Zhang and Junjing Guo and Junzhe Lin and Kaixiang Li and Lei Liu and Lei Xia and Liang Zhao and Liguo Tan and Liwen Huang and Liying Shi and Ming Li and Mingliang Li and Muhua Cheng and Na Wang and Qiaohui Chen and Qinglin He and Qiuyan Liang and Quan Sun and Ran Sun and Rui Wang and Shaoliang Pang and Shiliang Yang and Sitong Liu and Siqi Liu and Shuli Gao and Tiancheng Cao and Tianyu Wang and Weipeng Ming and Wenqing He and Xu Zhao and Xuelin Zhang and Xianfang Zeng and Xiaojia Liu and Xuan Yang and Yaqi Dai and Yanbo Yu and Yang Li and Yineng Deng and Yingming Wang and Yilei Wang and Yuanwei Lu and Yu Chen and Yu Luo and Yuchu Luo and Yuhe Yin and Yuheng Feng and Yuxiang Yang and Zecheng Tang and Zekai Zhang and Zidong Yang and Binxing Jiao and Jiansheng Chen and Jing Li and Shuchang Zhou and Xiangyu Zhang and Xinhao Zhang and Yibo Zhu and Heung-Yeung Shum and Daxin Jiang},
  year={2025},
  eprint={2502.10248},
  archivePrefix={arXiv},
  primaryClass={cs.CV},
  url={https://arxiv.org/abs/2502.10248},
}

@misc{step-audio-arxiv25,
  howpublished={arXiv},
  title={{Step-Audio}: Unified Understanding and Generation in Intelligent Speech Interaction},
  author={Ailin Huang and Boyong Wu and Bruce Wang and Chao Yan and Chen Hu and Chengli Feng and Fei Tian and Feiyu Shen and Jingbei Li and Mingrui Chen and Peng Liu and Ruihang Miao and Wang You and Xi Chen and Xuerui Yang and Yechang Huang and Yuxiang Zhang and Zheng Gong and Zixin Zhang and Hongyu Zhou and Jianjian Sun and Brian Li and Chengting Feng and Changyi Wan and Hanpeng Hu and Jianchang Wu and Jiangjie Zhen and Ranchen Ming and Song Yuan and Xuelin Zhang and Yu Zhou and Bingxin Li and Buyun Ma and Hongyuan Wang and Kang An and Wei Ji and Wen Li and Xuan Wen and Xiangwen Kong and Yuankai Ma and Yuanwei Liang and Yun Mou and Bahtiyar Ahmidi and Bin Wang and Bo Li and Changxin Miao and Chen Xu and Chenrun Wang and Dapeng Shi and Deshan Sun and Dingyuan Hu and Dula Sai and Enle Liu and Guanzhe Huang and Gulin Yan and Heng Wang and Haonan Jia and Haoyang Zhang and Jiahao Gong and Junjing Guo and Jiashuai Liu and Jiahong Liu and Jie Feng and Jie Wu and Jiaoren Wu and Jie Yang and Jinguo Wang and Jingyang Zhang and Junzhe Lin and Kaixiang Li and Lei Xia and Li Zhou and Liang Zhao and Longlong Gu and Mei Chen and Menglin Wu and Ming Li and Mingxiao Li and Mingliang Li and Mingyao Liang and Na Wang and Nie Hao and Qiling Wu and Qinyuan Tan and Ran Sun and Shuai Shuai and Shaoliang Pang and Shiliang Yang and Shuli Gao and Shanshan Yuan and Siqi Liu and Shihong Deng and Shilei Jiang and Sitong Liu and Tiancheng Cao and Tianyu Wang and Wenjin Deng and Wuxun Xie and Weipeng Ming and Wenqing He and Wen Sun and Xin Han and Xin Huang and Xiaomin Deng and Xiaojia Liu and Xin Wu and Xu Zhao and Yanan Wei and Yanbo Yu and Yang Cao and Yangguang Li and Yangzhen Ma and Yanming Xu and Yaoyu Wang and Yaqiang Shi and Yilei Wang and Yizhuang Zhou and Yinmin Zhong and Yang Zhang and Yaoben Wei and Yu Luo and Yuanwei Lu and Yuhe Yin and Yuchu Luo and Yuanhao Ding and Yuting Yan and Yaqi Dai and Yuxiang Yang and Zhe Xie and Zheng Ge and Zheng Sun and Zhewei Huang and Zhichao Chang and Zhisheng Guan and Zidong Yang and Zili Zhang and Binxing Jiao and Daxin Jiang and Heung-Yeung Shum and Jiansheng Chen and Jing Li and Shuchang Zhou and Xiangyu Zhang and Xinhao Zhang and Yibo Zhu},
  year={2025},
  eprint={2502.11946},
  archivePrefix={arXiv},
  primaryClass={cs.CL},
  url={https://arxiv.org/abs/2502.11946},
}

@inproceedings{spindle-asplos25,
  author = {Wang, Yujie and Zhu, Shenhan and Fu, Fangcheng and Miao, Xupeng and Zhang, Jie and Zhu, Juan and Hong, Fan and Li, Yong and Cui, Bin},
  title = {{Spindle}: Efficient Distributed Training of Multi-Task Large Models via Wavefront Scheduling},
  year = {2025},
  isbn = {9798400710797},
  publisher = {Association for Computing Machinery},
  address = {New York, NY, USA},
  url = {https://doi.org/10.1145/3676641.3715992},
  doi = {10.1145/3676641.3715992},
  booktitle = {Proceedings of the 30th ACM International Conference on Architectural Support for Programming Languages and Operating Systems, Volume 2},
  pages = {1139–1155},
  numpages = {17},
  location = {Rotterdam, Netherlands},
  series = {ASPLOS '25}
}

@misc{deepseek-v3-arxiv25,
      title={DeepSeek-V3 Technical Report},
      author={DeepSeek-AI and Aixin Liu and Bei Feng and Bing Xue and Bingxuan Wang and Bochao Wu and Chengda Lu and Chenggang Zhao and Chengqi Deng and Chenyu Zhang and Chong Ruan and Damai Dai and Daya Guo and Dejian Yang and Deli Chen and Dongjie Ji and Erhang Li and Fangyun Lin and Fucong Dai and Fuli Luo and Guangbo Hao and Guanting Chen and Guowei Li and H. Zhang and Han Bao and Hanwei Xu and Haocheng Wang and Haowei Zhang and Honghui Ding and Huajian Xin and Huazuo Gao and Hui Li and Hui Qu and J. L. Cai and Jian Liang and Jianzhong Guo and Jiaqi Ni and Jiashi Li and Jiawei Wang and Jin Chen and Jingchang Chen and Jingyang Yuan and Junjie Qiu and Junlong Li and Junxiao Song and Kai Dong and Kai Hu and Kaige Gao and Kang Guan and Kexin Huang and Kuai Yu and Lean Wang and Lecong Zhang and Lei Xu and Leyi Xia and Liang Zhao and Litong Wang and Liyue Zhang and Meng Li and Miaojun Wang and Mingchuan Zhang and Minghua Zhang and Minghui Tang and Mingming Li and Ning Tian and Panpan Huang and Peiyi Wang and Peng Zhang and Qiancheng Wang and Qihao Zhu and Qinyu Chen and Qiushi Du and R. J. Chen and R. L. Jin and Ruiqi Ge and Ruisong Zhang and Ruizhe Pan and Runji Wang and Runxin Xu and Ruoyu Zhang and Ruyi Chen and S. S. Li and Shanghao Lu and Shangyan Zhou and Shanhuang Chen and Shaoqing Wu and Shengfeng Ye and Shengfeng Ye and Shirong Ma and Shiyu Wang and Shuang Zhou and Shuiping Yu and Shunfeng Zhou and Shuting Pan and T. Wang and Tao Yun and Tian Pei and Tianyu Sun and W. L. Xiao and Wangding Zeng and Wanjia Zhao and Wei An and Wen Liu and Wenfeng Liang and Wenjun Gao and Wenqin Yu and Wentao Zhang and X. Q. Li and Xiangyue Jin and Xianzu Wang and Xiao Bi and Xiaodong Liu and Xiaohan Wang and Xiaojin Shen and Xiaokang Chen and Xiaokang Zhang and Xiaosha Chen and Xiaotao Nie and Xiaowen Sun and Xiaoxiang Wang and Xin Cheng and Xin Liu and Xin Xie and Xingchao Liu and Xingkai Yu and Xinnan Song and Xinxia Shan and Xinyi Zhou and Xinyu Yang and Xinyuan Li and Xuecheng Su and Xuheng Lin and Y. K. Li and Y. Q. Wang and Y. X. Wei and Y. X. Zhu and Yang Zhang and Yanhong Xu and Yanhong Xu and Yanping Huang and Yao Li and Yao Zhao and Yaofeng Sun and Yaohui Li and Yaohui Wang and Yi Yu and Yi Zheng and Yichao Zhang and Yifan Shi and Yiliang Xiong and Ying He and Ying Tang and Yishi Piao and Yisong Wang and Yixuan Tan and Yiyang Ma and Yiyuan Liu and Yongqiang Guo and Yu Wu and Yuan Ou and Yuchen Zhu and Yuduan Wang and Yue Gong and Yuheng Zou and Yujia He and Yukun Zha and Yunfan Xiong and Yunxian Ma and Yuting Yan and Yuxiang Luo and Yuxiang You and Yuxuan Liu and Yuyang Zhou and Z. F. Wu and Z. Z. Ren and Zehui Ren and Zhangli Sha and Zhe Fu and Zhean Xu and Zhen Huang and Zhen Zhang and Zhenda Xie and Zhengyan Zhang and Zhewen Hao and Zhibin Gou and Zhicheng Ma and Zhigang Yan and Zhihong Shao and Zhipeng Xu and Zhiyu Wu and Zhongyu Zhang and Zhuoshu Li and Zihui Gu and Zijia Zhu and Zijun Liu and Zilin Li and Ziwei Xie and Ziyang Song and Ziyi Gao and Zizheng Pan},
      year={2025},
      eprint={2412.19437},
      archivePrefix={arXiv},
      primaryClass={cs.CL},
      url={https://arxiv.org/abs/2412.19437},
}

@misc{qwen2-vl-arxiv24,
  howpublished={arXiv},
  title={{Qwen2-VL}: Enhancing Vision-Language Model's Perception of the World at Any Resolution},
  author={Peng Wang and Shuai Bai and Sinan Tan and Shijie Wang and Zhihao Fan and Jinze Bai and Keqin Chen and Xuejing Liu and Jialin Wang and Wenbin Ge and Yang Fan and Kai Dang and Mengfei Du and Xuancheng Ren and Rui Men and Dayiheng Liu and Chang Zhou and Jingren Zhou and Junyang Lin},
  year={2024},
  eprint={2409.12191},
  archivePrefix={arXiv},
  primaryClass={cs.CV},
  url={https://arxiv.org/abs/2409.12191},
}

@misc{qwen2.5-vl-arxiv25,
  howpublished={arXiv},
  title={{Qwen2.5-VL} Technical Report},
  author={Shuai Bai and Keqin Chen and Xuejing Liu and Jialin Wang and Wenbin Ge and Sibo Song and Kai Dang and Peng Wang and Shijie Wang and Jun Tang and Humen Zhong and Yuanzhi Zhu and Mingkun Yang and Zhaohai Li and Jianqiang Wan and Pengfei Wang and Wei Ding and Zheren Fu and Yiheng Xu and Jiabo Ye and Xi Zhang and Tianbao Xie and Zesen Cheng and Hang Zhang and Zhibo Yang and Haiyang Xu and Junyang Lin},
  year={2025},
  eprint={2502.13923},
  archivePrefix={arXiv},
  primaryClass={cs.CV},
  url={https://arxiv.org/abs/2502.13923},
}

@misc{gpt4o-arxiv24,
  title={{GPT-4o} System Card},
  author={OpenAI},
  year={2024},
  howpublished={\url{https://openai.com/index/gpt-4o-system-card/}},
}

@misc{gemini-3-2025,
  title={{Gemma 3}},
  author={DeepMind},
  year={2025},
  howpublished={\url{https://blog.google/technology/developers/gemma-3/}},
}

@inproceedings{transformer-nips17,
  author={Vaswani, Ashish and Shazeer, Noam and Parmar, Niki and Uszkoreit, Jakob and Jones, Llion and Gomez, Aidan N. and Kaiser, \L{}ukasz and Polosukhin, Illia},
  title={Attention is All You Need},
  booktitle = {Conference on Neural Information Processing Systems},
  year={2017},
  isbn={9781510860964},
  publisher={Curran Associates Inc.},
  address={Red Hook, NY, USA},
  pages={6000–6010},
  numpages={11},
  location={Long Beach, California, USA},
  series={NIPS'17},
}

@misc{modality-bridge-arxiv23,
  howpublished={arXiv},
  title={How to Bridge the Gap between Modalities: Survey on Multimodal Large Language Model},
  author={Shezheng Song and Xiaopeng Li and Shasha Li and Shan Zhao and Jie Yu and Jun Ma and Xiaoguang Mao and Weimin Zhang},
  year={2025},
  eprint={2311.07594},
  archivePrefix={arXiv},
  primaryClass={cs.CL},
  url={https://arxiv.org/abs/2311.07594},
}

@misc{hunyuan-dit-arxiv24,
  howpublished={arXiv},
  title={{Hunyuan-DiT}: A Powerful Multi-Resolution Diffusion Transformer with Fine-Grained Chinese Understanding},
  author={Zhimin Li and Jianwei Zhang and Qin Lin and Jiangfeng Xiong and Yanxin Long and Xinchi Deng and Yingfang Zhang and Xingchao Liu and Minbin Huang and Zedong Xiao and Dayou Chen and Jiajun He and Jiahao Li and Wenyue Li and Chen Zhang and Rongwei Quan and Jianxiang Lu and Jiabin Huang and Xiaoyan Yuan and Xiaoxiao Zheng and Yixuan Li and Jihong Zhang and Chao Zhang and Meng Chen and Jie Liu and Zheng Fang and Weiyan Wang and Jinbao Xue and Yangyu Tao and Jianchen Zhu and Kai Liu and Sihuan Lin and Yifu Sun and Yun Li and Dongdong Wang and Mingtao Chen and Zhichao Hu and Xiao Xiao and Yan Chen and Yuhong Liu and Wei Liu and Di Wang and Yong Yang and Jie Jiang and Qinglin Lu},
  year={2024},
  eprint={2405.08748},
  archivePrefix={arXiv},
  primaryClass={cs.CV},
  url={https://arxiv.org/abs/2405.08748},
}

@misc{dialog-gen-arxiv24,
  howpublished={arXiv},
  title={{DialogGen}: Multi-modal Interactive Dialogue System for Multi-turn Text-to-Image Generation},
  author={Minbin Huang and Yanxin Long and Xinchi Deng and Ruihang Chu and Jiangfeng Xiong and Xiaodan Liang and Hong Cheng and Qinglin Lu and Wei Liu},
  year={2024},
  eprint={2403.08857},
  archivePrefix={arXiv},
  primaryClass={cs.CV},
  url={https://arxiv.org/abs/2403.08857},
}

@misc{gpt4o-image-generation,
  title = {Introducing 4o Image Generation},
  author = {OpenAI},
  howpublished = {\url{https://openai.com/index/introducing-4o-image-generation/}},
  year         = {2025},
}

@misc{mlongdoc-arxiv24,
  howpublished={arXiv},
  title={{M-Longdoc}: A Benchmark For Multimodal Super-Long Document Understanding And A Retrieval-Aware Tuning Framework},
  author={Yew Ken Chia and Liying Cheng and Hou Pong Chan and Chaoqun Liu and Maojia Song and Sharifah Mahani Aljunied and Soujanya Poria and Lidong Bing},
  year={2024},
  eprint={2411.06176},
  archivePrefix={arXiv},
  primaryClass={cs.CL},
  url={https://arxiv.org/abs/2411.06176},
}

@inproceedings{dynapipe-eurosys24,
  author = {Jiang, Chenyu and Jia, Zhen and Zheng, Shuai and Wang, Yida and Wu, Chuan},
  title = {{DynaPipe}: Optimizing Multi-Task Training through Dynamic Pipelines},
  year = {2024},
  isbn = {9798400704376},
  publisher = {Association for Computing Machinery},
  address = {New York, NY, USA},
  url = {https://doi.org/10.1145/3627703.3629585},
  doi = {10.1145/3627703.3629585},
  booktitle = {Proceedings of the Nineteenth European Conference on Computer Systems},
  pages = {542–559},
  numpages = {18},
  location = {Athens, Greece},
  series = {EuroSys '24}
}

@misc{obelics-arxiv23,
  howpublished={arXiv},
  title={{OBELICS}: An Open Web-Scale Filtered Dataset of Interleaved Image-Text Documents},
  author={Hugo Laurençon and Lucile Saulnier and Léo Tronchon and Stas Bekman and Amanpreet Singh and Anton Lozhkov and Thomas Wang and Siddharth Karamcheti and Alexander M. Rush and Douwe Kiela and Matthieu Cord and Victor Sanh},
  year={2023},
  eprint={2306.16527},
  archivePrefix={arXiv},
  primaryClass={cs.IR},
  url={https://arxiv.org/abs/2306.16527},
}

@misc{mmc4-arxiv23,
  howpublished={arXiv},
  title={Multimodal {C4}: An Open, Billion-Scale Corpus of Images Interleaved with Text},
  author={Wanrong Zhu and Jack Hessel and Anas Awadalla and Samir Yitzhak Gadre and Jesse Dodge and Alex Fang and Youngjae Yu and Ludwig Schmidt and William Yang Wang and Yejin Choi},
  year={2023},
  eprint={2304.06939},
  archivePrefix={arXiv},
  primaryClass={cs.CV},
  url={https://arxiv.org/abs/2304.06939},
}

@misc{clip-arxiv21,
  howpublished={arXiv},
  title={Learning Transferable Visual Models From Natural Language Supervision},
  author={Alec Radford and Jong Wook Kim and Chris Hallacy and Aditya Ramesh and Gabriel Goh and Sandhini Agarwal and Girish Sastry and Amanda Askell and Pamela Mishkin and Jack Clark and Gretchen Krueger and Ilya Sutskever},
  year={2021},
  eprint={2103.00020},
  archivePrefix={arXiv},
  primaryClass={cs.CV},
  url={https://arxiv.org/abs/2103.00020},
}

@misc{coyo-700m,
  title         = {{COYO-700M}: Image-Text Pair Dataset},
  author        = {Byeon, Minwoo and Park, Beomhee and Kim, Haecheon and Lee, Sungjun and Baek, Woonhyuk and Kim, Saehoon},
  year          = {2022},
  howpublished  = {\url{https://github.com/kakaobrain/coyo-dataset}},
}

@misc{share-gpt4-video-arxiv24,
  howpublished={arXiv},
  title={{ShareGPT4Video}: Improving Video Understanding and Generation with Better Captions},
  author={Lin Chen and Xilin Wei and Jinsong Li and Xiaoyi Dong and Pan Zhang and Yuhang Zang and Zehui Chen and Haodong Duan and Bin Lin and Zhenyu Tang and Li Yuan and Yu Qiao and Dahua Lin and Feng Zhao and Jiaqi Wang},
  year={2024},
  eprint={2406.04325},
  archivePrefix={arXiv},
  primaryClass={cs.CV},
  url={https://arxiv.org/abs/2406.04325},
}

@misc{internvid-arxiv23,
  howpublished={arXiv},
  title={{InternVid}: A Large-Scale Video-Text Dataset for Multimodal Understanding and Generation},
  author={Yi Wang and Yinan He and Yizhuo Li and Kunchang Li and Jiashuo Yu and Xin Ma and Xinhao Li and Guo Chen and Xinyuan Chen and Yaohui Wang and Conghui He and Ping Luo and Ziwei Liu and Yali Wang and Limin Wang and Yu Qiao},
  year={2024},
  eprint={2307.06942},
  archivePrefix={arXiv},
  primaryClass={cs.CV},
  url={https://arxiv.org/abs/2307.06942},
}

@misc{mmtrail-arxiv24,
  howpublished={arXiv},
  title={{MMTrail}: A Multimodal Trailer Video Dataset with Language and Music Descriptions},
  author={Xiaowei Chi and Yatian Wang and Aosong Cheng and Pengjun Fang and Zeyue Tian and Yingqing He and Zhaoyang Liu and Xingqun Qi and Jiahao Pan and Rongyu Zhang and Mengfei Li and Ruibin Yuan and Yanbing Jiang and Wei Xue and Wenhan Luo and Qifeng Chen and Shanghang Zhang and Qifeng Liu and Yike Guo},
  year={2024},
  eprint={2407.20962},
  archivePrefix={arXiv},
  primaryClass={cs.CV},
  url={https://arxiv.org/abs/2407.20962},
}

@inproceedings{scienceqa-nips22,
  title={Learn to Explain: Multimodal Reasoning via Thought Chains for Science Question Answering},
  author={Lu, Pan and Mishra, Swaroop and Xia, Tony and Qiu, Liang and Chang, Kai-Wei and Zhu, Song-Chun and Tafjord, Oyvind and Clark, Peter and Ashwin Kalyan},
  booktitle={The 36th Conference on Neural Information Processing Systems (NeurIPS)},
  year={2022}
}

@inproceedings{msr-vtt-cvpr16,
  author={Xu, Jun and Mei, Tao and Yao, Ting and Rui, Yong},
  booktitle={2016 IEEE Conference on Computer Vision and Pattern Recognition (CVPR)},
  title={{MSR-VTT}: A Large Video Description Dataset for Bridging Video and Language},
  year={2016},
  volume={},
  number={},
  pages={5288-5296},
  doi={10.1109/CVPR.2016.571}
}

@inproceedings{recycle-sosp24,
  author = {Gandhi, Swapnil and Zhao, Mark and Skiadopoulos, Athinagoras and Kozyrakis, Christos},
  title = {{ReCycle}: Resilient Training of Large {DNNs} using Pipeline Adaptation},
  year = {2024},
  isbn = {9798400712517},
  publisher = {Association for Computing Machinery},
  address = {New York, NY, USA},
  url = {https://doi.org/10.1145/3694715.3695960},
  doi = {10.1145/3694715.3695960},
  booktitle = {Proceedings of the ACM SIGOPS 30th Symposium on Operating Systems Principles},
  pages = {211–228},
  numpages = {18},
  location = {Austin, TX, USA},
  series = {SOSP '24}
}

@misc{highs,
  title = {{HiGHS}: Linear Optimization Software},
  author = {HiGHS Team},
  howpublished = {\url{https://github.com/ERGO-Code/HiGHS}},
  year         = {2025},
}

@misc{gurobi,
  title = {{Gurobi}},
  author = {Gurobi Optimization},
  howpublished = {\url{https://www.gurobi.com/}},
  year         = {2025},
}

@inproceedings{simclr-icml20,
  author = {Chen, Ting and Kornblith, Simon and Norouzi, Mohammad and Hinton, Geoffrey},
  title = {A Simple Framework for Contrastive Learning of Visual Representations},
  year = {2020},
  publisher = {JMLR.org},
  booktitle = {Proceedings of the 37th International Conference on Machine Learning},
  articleno = {149},
  numpages = {11},
  series = {ICML'20},
  address = {Virtual conference}
}

@misc{fsdp-arxiv23,
  howpublished={arXiv},
  title={{PyTorch} {FSDP}: Experiences on Scaling Fully Sharded Data Parallel},
  author={Yanli Zhao and Andrew Gu and Rohan Varma and Liang Luo and Chien-Chin Huang and Min Xu and Less Wright and Hamid Shojanazeri and Myle Ott and Sam Shleifer and Alban Desmaison and Can Balioglu and Pritam Damania and Bernard Nguyen and Geeta Chauhan and Yuchen Hao and Ajit Mathews and Shen Li},
  year={2023},
  eprint={2304.11277},
  archivePrefix={arXiv},
  primaryClass={cs.DC},
  url={https://arxiv.org/abs/2304.11277},
}

@article{multi-choice-knapsack,
  author = {Armstrong, R. D. and Kung, D. S. and Sinha, P. and Zoltners, A. A.},
  title = {A Computational Study of a Multiple-Choice Knapsack Algorithm},
  year = {1983},
  issue_date = {June 1983},
  publisher = {Association for Computing Machinery},
  address = {New York, NY, USA},
  volume = {9},
  number = {2},
  issn = {0098-3500},
  url = {https://doi.org/10.1145/357456.357458},
  doi = {10.1145/357456.357458},
  journal = {ACM Trans. Math. Softw.},
  month = jun,
  pages = {184–198},
  numpages = {15}
}

@inproceedings{iverson-bracket,
  author = {Iverson, Kenneth E.},
  title = {A Programming Language},
  year = {1962},
  isbn = {9781450378758},
  publisher = {Association for Computing Machinery},
  address = {New York, NY, USA},
  url = {https://doi.org/10.1145/1460833.1460872},
  doi = {10.1145/1460833.1460872},
  booktitle = {Proceedings of the May 1-3, 1962, Spring Joint Computer Conference},
  pages = {345–351},
  numpages = {7},
  location = {San Francisco, California},
  series = {AIEE-IRE '62 (Spring)}
}

@inproceedings{distmm-nsdi24,
  author = {Jun Huang and Zhen Zhang and Shuai Zheng and Feng Qin and Yida Wang},
  title = {{DISTMM}: Accelerating Distributed Multimodal Model Training},
  booktitle = {21st USENIX Symposium on Networked Systems Design and Implementation (NSDI 24)},
  year = {2024},
  isbn = {978-1-939133-39-7},
  address = {Santa Clara, CA},
  pages = {1157--1171},
  url = {https://www.usenix.org/conference/nsdi24/presentation/huang},
  publisher = {USENIX Association},
  month = apr
}

@misc{orch-mllm-arxiv25,
  howpublished={arXiv},
  title={Orchestrate Multimodal Data with Batch Post-Balancing to Accelerate Multimodal Large Language Model Training},
  author={Yijie Zheng and Bangjun Xiao and Lei Shi and Xiaoyang Li and Faming Wu and Tianyu Li and Xuefeng Xiao and Yang Zhang and Yuxuan Wang and Shouda Liu},
  year={2025},
  eprint={2503.23830},
  archivePrefix={arXiv},
  primaryClass={cs.DC},
  url={https://arxiv.org/abs/2503.23830},
}

@misc{cornstarch-arxiv25,
  howpublished={arXiv},
  title={{Cornstarch}: Distributed Multimodal Training Must Be Multimodality-Aware},
  author={Insu Jang and Runyu Lu and Nikhil Bansal and Ang Chen and Mosharaf Chowdhury},
  year={2025},
  eprint={2503.11367},
  archivePrefix={arXiv},
  primaryClass={cs.DC},
  url={https://arxiv.org/abs/2503.11367},
}

@INPROCEEDINGS {byte-transformer-ipdps23,
author = { Zhai, Yujia and Jiang, Chengquan and Wang, Leyuan and Jia, Xiaoying and Zhang, Shang and Chen, Zizhong and Liu, Xin and Zhu, Yibo },
booktitle = { 2023 IEEE International Parallel and Distributed Processing Symposium (IPDPS) },
title = {{ByteTransformer}: A High-Performance Transformer Boosted for Variable-Length Inputs},
year = {2023},
volume = {},
ISSN = {},
pages = {344-355},
doi = {10.1109/IPDPS54959.2023.00042},
url = {https://doi.ieeecomputersociety.org/10.1109/IPDPS54959.2023.00042},
publisher = {IEEE Computer Society},
address = {Los Alamitos, CA, USA},
month =May
}

@misc{sppo-arxiv25,
      title={{SPPO}: Efficient Long-Sequence {LLM} Training via Adaptive Sequence Pipeline Parallel Offloading},
      author={Qiaoling Chen and Shenggui Li and Wei Gao and Peng Sun and Yonggang Wen and Tianwei Zhang},
      year={2025},
      eprint={2503.10377},
      archivePrefix={arXiv},
      primaryClass={cs.DC},
      url={https://arxiv.org/abs/2503.10377},
}

@inproceedings{graph-pipe-asplos25,
author = {Jeon, Byungsoo and Wu, Mengdi and Cao, Shiyi and Kim, Sunghyun and Park, Sunghyun and Aggarwal, Neeraj and Unger, Colin and Arfeen, Daiyaan and Liao, Peiyuan and Miao, Xupeng and Alizadeh, Mohammad and Ganger, Gregory R. and Chen, Tianqi and Jia, Zhihao},
title = {{GraphPipe}: Improving Performance and Scalability of {DNN} Training with Graph Pipeline Parallelism},
year = {2025},
isbn = {9798400706981},
publisher = {Association for Computing Machinery},
address = {New York, NY, USA},
url = {https://doi.org/10.1145/3669940.3707220},
doi = {10.1145/3669940.3707220},
booktitle = {Proceedings of the 30th ACM International Conference on Architectural Support for Programming Languages and Operating Systems, Volume 1},
pages = {557–571},
numpages = {15},
location = {Rotterdam, Netherlands},
series = {ASPLOS '25}
}

@inproceedings{chimera-sc21,
author = {Li, Shigang and Hoefler, Torsten},
title = {{Chimera}: Efficiently Training Large-Scale Neural Networks with Bidirectional Pipelines},
year = {2021},
isbn = {9781450384421},
publisher = {Association for Computing Machinery},
address = {New York, NY, USA},
url = {https://doi.org/10.1145/3458817.3476145},
doi = {10.1145/3458817.3476145},
booktitle = {Proceedings of the International Conference for High Performance Computing, Networking, Storage and Analysis},
articleno = {27},
numpages = {14},
location = {St. Louis, Missouri},
series = {SC '21}
}

@inproceedings{hanayo-sc23,
author = {Liu, Ziming and Cheng, Shenggan and Zhou, Haotian and You, Yang},
title = {{Hanayo}: Harnessing Wave-Like Pipeline Parallelism for Enhanced Large Model Training Efficiency},
year = {2023},
isbn = {9798400701092},
publisher = {Association for Computing Machinery},
address = {New York, NY, USA},
url = {https://doi.org/10.1145/3581784.3607073},
doi = {10.1145/3581784.3607073},
booktitle = {Proceedings of the International Conference for High Performance Computing, Networking, Storage and Analysis},
articleno = {56},
numpages = {13},
location = {Denver, CO, USA},
series = {SC '23}
}

@inproceedings{zero-bubble-iclr24,
title={Zero Bubble Pipeline Parallelism},
author={Penghui Qi and Xinyi Wan and Guangxing Huang and Min Lin},
booktitle={The Twelfth International Conference on Learning Representations},
year={2024},
url={https://openreview.net/forum?id=tuzTN0eIO5}
}


\end{document}